\documentclass[12pt,usenatbib,referee]{article}

\usepackage[mathscr]{eucal}
\usepackage{bbm,mathtools,amsfonts,amsmath,amssymb,graphicx,bm,latexsym}
\usepackage[]{natbib}
\bibpunct{(}{)}{;}{a}{}{,}
\usepackage{tikz}

\usetikzlibrary{bayesnet}
\newcommand{\obs}{latent}
\newcommand{\lat}{obs}

\usepackage[tikz]{bclogo}

\newcommand{\bt}{\begin{bclogo}[couleur={rgb:orange,0;yellow,0;white,1},arrondi=0.1,logo=\bcplume,ombre=true]}
\newcommand{\et}{\end{bclogo}\s}
\newcommand{\btt}{\begin{box}}
\newcommand{\ett}{\end{box}}

\newcommand{\btheorem}{\begin{bclogo}[couleur={rgb:orange,0;yellow,0;white,1},arrondi=0.1,logo=\bcplume,ombre=true]{Theorem}}
\newcommand{\ettheorem}{\end{bclogo}}

\newcommand{\bst}{\begin{bclogo}[couleur={rgb:orange,1;yellow,1;white,0.5},arrondi=0.1,logo=\bcpanchant]}
\newcommand{\est}{\end{bclogo}}

\newcommand{\benum}{\begin{enumerate}}
\newcommand{\eenum}{\end{enumerate}}

\newcommand{\bq}{\begin{quote}\em}
\newcommand{\eq}{\end{quote}}
\newcommand{\bbq}{\begin{quote}\bf\em}
\newcommand{\ebq}{\end{quote}}

\newcommand{\ind}{\msim\limits^{\mbox{\tiny ind}}}
\newcommand{\iid}{\msim\limits^{\mbox{\tiny iid}}}

\renewcommand{\=}{&=&}

\newcommand{\mR}{\mathbb{R}}

\newcommand{\mbE}{\mathbb{E}}

\newcommand{\mbP}{\mathbb{P}}

\newcommand{\hide}[1]{}
\newcommand{\ba}{\begin{array}{llllllllll}}
\newcommand{\ea}{\end{array}}
\newcommand{\bea}{\begin{equation}\begin{array}{llllllllll}\nonumber}
\newcommand{\eea}{\end{array}\end{equation}}
\newcommand{\be}{\begin{equation}\begin{array}{lllllllllllllllll}\nonumber}
\newcommand{\beno}{\begin{equation}\begin{array}{lllllllllllll}\nonumber}
\newcommand{\ee}{\end{array}\end{equation}}
\newcommand{\bel}{\begin{equation}\begin{array}{lllllllllllll}\nonumber}
\newcommand{\eel}{\Box\end{array}\end{equation}}
\newcommand{\bi}{\begin{itemize}}
\newcommand{\ei}{\end{itemize}}
\newcommand{\ben}{\begin{enumerate}}
\newcommand{\een}{\end{enumerate}}
\newcommand{\alert}{\textcolor{red}}
\newcommand{\dis}{\displaystyle}
\newcommand{\dsum}{\displaystyle\sum\limits}

\newcommand{\dd}{\mathop{\mbox{d}}\nolimits}

\newcommand{\mM}{\mbox{$\mathbb{M}$}}

\newcommand{\mA}{\mathscr{A}}

\newcommand{\s}{\vspace{0.25cm}}
\newcommand{\bx}{\bm{x}}

\newcommand{\baa}{\bm{a}}
\newcommand{\bA}{\bm{A}}

\newcommand{\bT}{\bm{T}}
\newcommand{\bX}{\bm{X}}
\newcommand{\mX}{\mathbb{X}}

\newcommand{\mY}{\mathscr{Y}}

\newcommand{\bY}{\bm{Y}}

\newcommand{\bbY}{\mathscr{Y}}
\newcommand{\by}{\bm{y}}

\newcommand{\bz}{\bm{z}}
\newcommand{\bZ}{\bm{Z}}

\newcommand{\bbeta}{\mbox{\boldmath$\beta$}}
\newcommand{\bgamma}{\mbox{\boldmath$\gamma$}}

\newcommand{\bmu}{\mbox{\boldmath$\mu$}}

\newcommand{\bSigma}{\mbox{\boldmath$\Sigma$}}
\newcommand{\bTheta}{\mbox{\boldmath$\Theta$}}
\newcommand{\bta}{\boldsymbol{\eta}}
\newcommand{\btheta}{\boldsymbol{\theta}}

\newcommand{\bvartheta}{\boldsymbol{\vartheta}}
\newcommand{\bpi}{\mbox{\boldmath$\pi$}}

\newcommand{\msim}{\mathop{\rm \sim}}

\newcommand{\bI}{\bm{I}}
\newcommand{\bR}{\bm{R}}
\newcommand{\bE}{\bm{E}}

\newcounter{comment}
\newenvironment{comment}[1][]{\refstepcounter{comment}\vspace{0.25cm}\par\medskip\noindent%
\textbf{Comment~\thecomment #1}:\vspace{0.25cm}\\ \rmfamily}{\medskip}
\newcommand{\bc}{\begin{comment}\em}
\newcommand{\ec}{\end{comment}}

\newcounter{proposition}
\newenvironment{proposition}[1][]{\refstepcounter{proposition}\par\medskip\indent%
\textbf{Proposition~\theproposition #1}. \rmfamily}{\medskip}

\newcounter{com}

\renewcommand{\alert}{\textcolor{black}}

\usepackage[letterpaper, portrait, margin=1in]{geometry}

\newcommand{\longtitle}{A Semiparametric Bayesian Approach to Epidemics,\linebreak
with Application to the Spread of the Coronavirus MERS in South Korea in 2015}

\title{\longtitle}

\author{Michael Schweinberger\\
Rice University
\and
Rashmi P.\ Bomiriya\\
Penn State University\\
\and
Sergii Babkin\\
Rice University
}

\date{}

\begin{document}

\maketitle

\vspace{-1cm}

\begin{abstract}
We consider incomplete observations of stochastic processes governing the spread of infectious diseases through finite populations by way of contact.
We propose a flexible semiparametric modeling framework with at least three advantages.
First,
it enables researchers to study the structure of a population contact network and its impact on the spread of infectious diseases.
Second,
it can accommodate short- and long-tailed degree distributions and detect potential superspreaders,
who represent an important public health concern.
Third,
it addresses the important issue of incomplete data.
Starting from first principles,
we show when the incomplete-data generating process is ignorable for the purpose of Bayesian inference for the parameters of the population model.
We demonstrate the semiparametric modeling framework by simulations and an application to the partially observed MERS epidemic in South Korea in 2015.
We conclude with an extended discussion of open questions and directions for future research.

{\em Keywords:} Contact networks; Network sampling; Link-tracing; Missing data.

\end{abstract}

\tableofcontents

\section{Motivation}
\label{introduction}

\alert{The spread of infectious diseases through populations by way of contact represents an important public health concern.
The spread of COVID-19 and other viruses underscores the importance of understanding how infectious diseases spread by way of contact and how the spread of infectious diseases can be curbed.
Indeed,
COVID-19 is not the first virus to spread around the globe,
and it will not be the last:
Epidemics have been documented since at least the Middle Ages (e.g., the plague) and are primed to become more frequent rather than less frequent in the interconnected world of the twenty-first century (as the recent spread of COVID-19, MERS, SARS, Ebola, and Zika demonstrates).}

\subsection{Advantages of network-based approaches to epidemics}

We follow a network-based approach to modeling the spread of infectious diseases.\linebreak
A network-based approach is appealing,
because the network of contacts in a population determines how infectious diseases can spread (e.g., \citeauthor{KeEa05} \citeyear{KeEa05}; \citeauthor{Daetal11} \citeyear{Daetal11}; \citeauthor*{WeBaHu11} \citeyear{WeBaHu11}).
One of the advantages of a network-based approach is that heterogeneity in the number of contacts can be captured \citep[e.g.,][]{Daetal11} along with other features of population contact networks \citep[e.g.,][]{WeBaHu11}.
Indeed,
conventional models of epidemics---including classic and lattice-based Susceptible-Infectious-Recovered (SIR) and Susceptible-Exposed-Infectious-Recovered (SEIR) models \citep[e.g.,][]{AnBr00,Daetal11}---can be considered to be degenerate versions of network-based models,
in the sense that such models postulate that with probability $1$ the population contact network is of a known form:
e.g.,
with probability $1$ each population member is in contact with every other population member.
Worse,
the postulated form of the population contact network may not resemble real-world contact networks.
A second advantage is that a network-based approach helps study the structure of a population contact network and its generating mechanism,
helping generalize findings to similar populations.

\subsection{Shortcomings of existing network-based approaches}

Motivated by the shortcomings of classic and lattice-based SIR and SEIR models of epidemics,
\citet{BrNe02},
\citet*{GrWeHu10,GrWeHu11},
\citet{epinet.jss},
\citet*{Buetal21},
and others have explored a network-based approach to epidemics.
While a network-based approach is more appealing than classic and lattice-based SIR and SEIR models,
existing network-based models of epidemics have shortcomings.
As we discuss in Section \ref{sec:shortcomings}, 
one of the more important shortcomings is that existing network-based approaches are either not flexible models of degree distributions or induce short-tailed degree distributions,
that is,
the population does not contain population members who have far more contacts than the bulk of the population members.
Short-tailed degree distributions are problematic,
because degree distributions of real-world contact networks are thought to be long-tailed \citep*[e.g.,][]{La94,AlBa02,JoHa04,JoHa03,HaJo04} and the population members in the upper tail of the degree distribution represent an important public health concern:
Population members with many contacts can infect many others and are hence potential superspreaders.
Indeed,
there is circumstantial evidence to suggest that superspreaders have played a role in the SARS epidemic in 2002--2003, the MERS epidemic in 2015, and the ongoing COVID-19 pandemic.

\subsection{Proposed network-based approach}

We introduce a flexible semiparametric modeling framework based on infinite mixture models and Dirichlet process priors \citep{Fe73,Teh2007},
with at least three advantages.
First,
it shares with existing network-based approaches the advantage that it enables researchers to study the structure of a population contact network and its impact on the spread of infectious diseases.
Second,
in contrast to existing network-based approaches,
it is a flexible model of short- and long-tailed degree distributions and can detect potential superspreaders.
Third,
it addresses the important issue of incomplete data and can deal with a wide range of missing data and sample data.
Starting from first principles,
we show when the incomplete-data generating process is ignorable for the purpose of Bayesian inference for the parameters of the population model.
In addition,
we stress the importance of collecting network data with a view to reducing the posterior uncertainty about the population contact network and its generating mechanism along with possible sources of infections.
We demonstrate the proposed framework by simulations and an application to the partially observed MERS epidemic in South Korea in 2015 \citep{Ki15}.
The MERS epidemic was driven by the coronavirus MERS,
which is related to the coronaviruses SARS and COVID-19.
We detect three to five potential superspreaders,
who may have had a great impact on the outcome of the outbreak.

\subsection{Goal: superpopulation inference for finite populations}

\alert{The proposed semiparametric modeling framework,
based on infinite mixture distributions and Dirichlet process priors \citep{Fe73,Teh2007},
extends to infinite populations.
That said,
we assume that the number of population members $N$ is finite and embrace a superpopulation approach to statistical inference along the lines of \citet{hartley1975super} and \citet*{ScKrBu17},
motivated by applications.

The assumption of finite $N$ is motivated by the fact that in epidemiological applications the number of population members $N$ cannot be infinite.
For example,
when the population of interest consists of all animals or all humans on earth,
the size of the population is bounded above by real-world constraints such as geography and the scarcity of natural resources:
Planet earth cannot host infinite populations of animals or humans.

Since the population of interest is finite,
the natural objective of statistical inference is to learn the stochastic process that generated the population contact network and allows infectious diseases to spread through the population of interest,
with a view to understanding and predicting epidemics in the population of interest and similar populations.
In other words,
it is natural to embrace a superpopulation approach to statistical inference,
as discussed by \citet{hartley1975super} and \citet{ScKrBu17}.
The properties of statistical procedures for superpopulation inference can be understood by developing a non-asymptotic statistical theory that relies on concentration inequalities and other non-asymptotic tools that have been embraced in high-dimensional statistics \citep[see, e.g.,][]{Wa19}.
We are not aware of non-asymptotic statistical theory for stochastic models of epidemics,
although there are asymptotic results in probability theory \citep*[e.g.,][]{Re95,BrLiTu11,BaRe13,PaPa20,Ba21} and statistical theory \citep[e.g.,][]{Br98,Br01a} based on $N \to \infty$ asymptotics.
Developing non-asymptotic statistical theory for stochastic models of epidemics constitutes an interesting direction for future research,
but is beyond the scope of our paper.
}

\subsection{Structure of the paper}

The remainder of the paper is structured as follows.
A network-based stochastic model of epidemics is reviewed in Section \ref{parametric}.
We discuss shortcomings of parametric population models in Section \ref{sec:shortcomings} and introduce semiparametric population models in Section \ref{population.model}.
In Section \ref{importance},
we argue that collecting complete data is all but impossible,
and stress the importance of collecting network data.
Principled Bayesian inference based on incomplete data is discussed in Section \ref{bay}.
We present simulation results in Section \ref{simulation.study} and an application to the partially observed MERS epidemic in South Korea in Section \ref{application}.
We conclude with an extended discussion of open questions and directions for future research in Section \ref{discussion}.

\section{A network-based stochastic model of epidemics}
\label{parametric}

We introduce a network-based stochastic model of epidemics.
We first describe a generic data-generating process in Section \ref{sec:data.generating.process} and then review parametric population models in Section \ref{likelihood.function}.

\subsection{Data-generating process}
\label{sec:data.generating.process}

\begin{figure}[t]
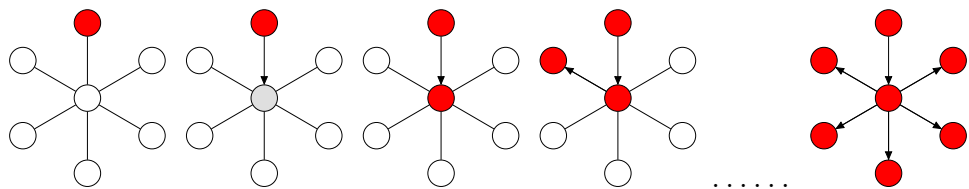

\center
\caption{\label{data.generating.process}Data-generating process: 
Conditional on contacts (undirected lines) among population members (circles), 
infectious population members (red) spread an infectious disease by contact (directed lines) to susceptible population members (white), which are exposed (gray) before turning infectious (red).}\s
\scalebox{0.5}{
      \tikz{ %
        \node[\obs] (x3) {} ; %
        \node[\obs, left=of x3, yshift=1cm] (x1) {} ; %
        \node[\obs, left=of x3, yshift=-1cm] (x2) {} ; %
        \node[\obs, right=of x3, yshift=1cm] (x4) {} ; %
        \node[\obs, right=of x3, yshift=-1cm] (x5) {} ; %
        \node[\obs, yshift=-2cm] (x6) {} ; %
        \node[inf, yshift=2cm] (x7) {} ; %
        \edge[-] {x1,x2,x4,x5,x6,x7} {x3} ; %
      }
}
\scalebox{0.5}{
      \tikz{ %
        \node[\lat] (x3) {} ; %
        \node[\obs, left=of x3, yshift=1cm] (x1) {} ; %
        \node[\obs, left=of x3, yshift=-1cm] (x2) {} ; %
        \node[\obs, right=of x3, yshift=1cm] (x4) {} ; %
        \node[\obs, right=of x3, yshift=-1cm] (x5) {} ; %
        \node[\obs, yshift=-2cm] (x6) {} ; %
        \node[inf, yshift=2cm] (x7) {} ; %
        \edge[-] {x1,x2,x4,x5,x6} {x3} ; %
        \edge[] {x7} {x3} ; %
      }
}
\scalebox{0.5}{
      \tikz{ %
        \node[inf] (x3) {} ; %
        \node[\obs, left=of x3, yshift=1cm] (x1) {} ; %
        \node[\obs, left=of x3, yshift=-1cm] (x2) {} ; %
        \node[\obs, right=of x3, yshift=1cm] (x4) {} ; %
        \node[\obs, right=of x3, yshift=-1cm] (x5) {} ; %
        \node[\obs, yshift=-2cm] (x6) {} ; %
        \node[inf, yshift=2cm] (x7) {} ; %
        \edge[-] {x1,x2,x4,x5,x6} {x3} ; %
        \edge[] {x7} {x3} ; %
      }
}
\scalebox{0.5}{
      \tikz{ %
        \node[inf] (x3) {} ; %
        \node[inf, left=of x3, yshift=1cm] (x1) {} ; %
        \node[\obs, left=of x3, yshift=-1cm] (x2) {} ; %
        \node[\obs, right=of x3, yshift=1cm] (x4) {} ; %
        \node[\obs, right=of x3, yshift=-1cm] (x5) {} ; %
        \node[\obs, yshift=-2cm] (x6) {} ; %
        \node[inf, yshift=2cm] (x7) {} ; %
        \edge[-] {x1,x2,x4,x5,x6} {x3} ; %
        \edge[] {x7} {x3} ; %
        \edge[] {x3} {x1} ; %
      }
}
$\dots\dots$
\scalebox{0.5}{
      \tikz{ %
        \node[inf] (x3) {} ; %
        \node[inf, left=of x3, yshift=1cm] (x1) {} ; %
        \node[inf, left=of x3, yshift=-1cm] (x2) {} ; %
        \node[inf, right=of x3, yshift=1cm] (x4) {} ; %
        \node[inf, right=of x3, yshift=-1cm] (x5) {} ; %
        \node[inf, yshift=-2cm] (x6) {} ; %
        \node[inf, yshift=2cm] (x7) {} ; %
        \edge[-] {x1,x2,x4,x5,x6} {x3} ; %
        \edge[] {x7} {x3} ; %
        \edge[] {x3} {x1,x2,x4,x5,x6} ; %
      }
}
\end{figure}

We consider a population with $N < \infty$ population members,
who may be connected by contacts.
In the simplest case,
contacts among population members are time-invariant and are either absent or present.
We discuss in Section \ref{temporal} possible extensions to time-evolving population contact networks.

A basic data-generating process is shown in Figure \ref{data.generating.process} and can be described as follows:
\bi
\item Generate a population contact network.
\item Conditional on the population contact network, 
generate an epidemic.
\ei
The population contact network is generated by a random graph model as described in Section \ref{likelihood.function} (parametric) and Section \ref{population.model} (semiparametric).
Conditional on the population contact network,
an infectious disease spreads through the population by way of contact,
governed by a continuous-time stochastic process such as the SIR and SEIR model \citep{AnBr00,BrNe02,GrWeHu10,GrWeHu11}.
We focus on the network-based SEIR model,
which can be sketched as follows \citep{BrNe02,GrWeHu10,GrWeHu11}.
In the simplest case,
the initial state of the stochastic process consists of a population with one infected population member and $N - 1$ susceptible population members.
Infected population members pass through three states:
the exposed state;
the infectious state;
and the removed state.
In the exposed state,
population members are infected, but cannot infect others.
In the infectious state, 
population members can infect susceptible population members by way of contact,
with transmissions being independent across contacts.
In the final state---the removed state---population members have either recovered and are immune to re-infection or have died,
and hence cannot infect others.
The epidemic continues until all infected population members are removed from the population.
All of the described events---the event that an infectious population member infects a susceptible population member and the transition of an infected population member from the exposed to the infectious state and from the infectious state to the removed state---are independent and occur at random times.
The waiting times until these events occur follow Exponential or Gamma distributions.
More specific assumptions about the distributions of waiting times and the population contact network are detailed in Section \ref{likelihood.function} and in the monographs of \citet{AnBr00} and \citet{epibook}.

\subsection{Parametric population models}
\label{likelihood.function}

Consider an epidemic that started at time $0$ and ceased by time $0 < t < \infty$.
We assume that the identities of infected population members are known and denoted by $1, \dots, M$,
where $M \in \{1, \dots, N\}$.
The population contact network is represented by $\bY = \{Y_{i,j}\}_{i < j}^N \in \mY = \{0, 1\}^{\binom{N}{2}}$,
where $Y_{i,j} = 1$ if population members $i$ and $j$ are in contact during the epidemic and $Y_{i,j} = 0$ otherwise.
Since contacts are undirected and self-contacts are meaningless,
we assume that $Y_{i,j} = Y_{j,i}$ and $Y_{i,i} = 0$ hold with probability 1.
The transmissions are denoted by $\bT = \{T_{i,j}\}_{i \neq j}^N$,
where $T_{i,j} = 1$ if $i$ infects $j$ and $T_{i,j} = 0$ otherwise.
Observe that $Y_{i,j} = 0$ implies $T_{i,j} = T_{j,i} = 0$ whereas $T_{i,j} = T_{j,i} = 1$ implies $Y_{i,j} = 1$ with probability 1.
The starting times of the exposure, infectious, and removal periods of infected population members are denoted by
$\bE = \{E_i\}_{i=1}^N \in \mR_+^N$,\,
$\bI = \{I_i\}_{i=1}^N \in \mR_+^N$,\,
and $\bR = \{R_i\}_{i=1}^N \in \mR_+^N$,
respectively,
where $\mR_+ = (0,\, \infty)$ and $E_i < I_i < R_i$ holds with probability 1;
note that $E_i,\, I_i,\, R_i$ are undefined if population member $i$ is not infected.
We write $\bX = (\bE, \bI, \bR, \bT)$ and, 
in a mild abuse of language,
we refer to $\bX$ as an epidemic. 

The complete-data likelihood function, 
given complete observations $\bx$ and $\by$ of the epidemic $\bX$ and the population contact network $\bY$,
is of the form
\bea
\label{eq:likelihood}
L(\bta,\, \btheta;\; \bx,\, \by) 
&\propto& L(\bta_E;\, \bx)\; L(\bta_I;\, \bx)\; L(\beta;\, \bx,\, \by)\; L(\btheta;\, \by),
\eea
where $\bta = (\bta_E,\, \bta_I,\, \beta) \in \Omega_{\bta} \subseteq \mR^{d_1}$ ($d_1 \geq 1$) and $\btheta \in \Omega_{\btheta} \subseteq \mR^{d_2}$ ($d_2 \geq 1$) are the parameter vectors of the population model generating the epidemic $\bX$ and the population contact network $\bY$,
respectively.
We describe each component of the complete-data likelihood function in turn,
along with its parameters.
\hide{
$\beta$ is the parameter governing the rate of infection;
and $\bta_E$ and $\bta_I$ are the parameter vectors governing the length of time spent in the exposed and infectious state,
respectively.
}

The components $L(\bta_E;\, \bx)$ and $L(\bta_I;\, \bx)$ of the likelihood function are of the form
\bea
L(\bta_E;\, \bx) 
&\propto& \dis\prod_{i=1}^M p(I_i - E_i \mid E_i,\, \bta_E)
\eea
\bea
L(\bta_I;\, \bx)
&\propto& \dis\prod_{i=1}^M p(R_i - I_i \mid I_i,\, \bta_I),
\eea
where $p(.\mid E_i,\, \bta_E)$ and $p(.\mid I_i,\, \bta_I)$ are densities with suitable support parameterized by $\bta_E$ and $\bta_I$,
respectively:
e.g.,
the densities may be Gamma densities,
with $\bta_E \in \mR_+ \times \mR_+$ and $\bta_I \in \mR_+ \times \mR_+$ being scale and shape parameters of Gamma densities.

Under the assumption that the waiting times until infectious population members infect susceptible population members are independent Exponential$(\beta)$ random variables with rate of infection $\beta \in \mR_+$,
the component $L(\beta;\, \bx,\, \by)$ of the likelihood function is given by
\bea
L(\beta;\, \bx,\, \by)
&\propto& \beta^{M-1}\, \exp(-\beta\; a(\bx,\, \by)),
\eea
where $a(.) > 0$ is defined by
\bea
\label{A}
a(\bx,\, \by) = \dsum_{i=1}^M \dsum_{j=1}^M y_{i,j}\, 1_{I_i < E_j} \max(\min(E_j, R_i) - I_i,\, 0) + \dsum_{i=1}^M d_i(\bx,\, \by)\, (R_i - I_i).
\eea
Here, 
$1_{I_i < E_j}$ is $1$ in the event of $I_i < E_j$ and is $0$ otherwise, 
and $d_i(\bx,\, \by)$ is the number of non-infected population members in contact with population member $i$.
The function $a(\bx,\, \by)$ was derived in \citet{BrNe02} and \citet{GrWeHu10}.

To represent existing (parametric) and proposed (semiparametric) population models along with possible extensions in a unifying framework,
it is convenient to parameterize the distribution of the population contact network $\bY$ in exponential-family form \citep{Su19}.
The component $L(\btheta;\, \by)$ of the likelihood function can then be written as
\bea
\label{general}
L(\btheta;\, \by) 
&\propto& \exp\left(\btheta^\top s(\by) - \psi(\btheta)\right), & \by \in \bbY,
\eea
where $\btheta$ is a $d_2$-vector of natural parameters, 
$s(\by)$ is a $d_2$-vector of sufficient statistics,
and 
\bea
\label{psi}
\psi(\btheta)
\= \log \dis\sum_{\by \in \mY} \exp\left(\btheta^\top s(\by)\right), & \btheta \in \Omega_{\btheta} = \{\btheta \in \mR^{d_2}: \psi(\btheta) < \infty\} = \mR^{d_2}.
\eea
\citet{BrNe02} and \citet{GrWeHu10} assumed that contacts are independent Bernoulli$(\mu)$ ($\mu \in (0, 1)$) random variables,
which is equivalent to the one-parameter exponential family with natural parameter $\theta = \mbox{logit}(\mu) \in \mR$ and sufficient statistic $s(\by) = \sum_{i<j}^N y_{i,j}$.
\citet{GrWeHu11} extended the exponential-family framework to include predictors of contacts by assuming that  contacts are independent Bernoulli$(\mu_{i,j})$ ($\mu_{i,j} \in (0, 1)$) random variables with $\mbox{logit}(\mu_{i,j}) = \sum_{k=1}^{d_2} \theta_k\, s_{i,j,k}(y_{i,j}) \in \mR$,
that is,
the log odds of the probability of a contact between two population members is a weighted sum of functions $s_{i,j,k}(y_{i,j})$ of covariates and $y_{i,j}$,
weighted by $\theta_k \in \mR$ ($k = 1, \dots, d_2$).
A specific example is given by $s_{i,j,1}(y_{i,j}) = y_{i,j}$ and $s_{i,j,2}(y_{i,j}) = c_{i,j}\, y_{i,j}$,
where $c_{i,j} \in \{0, 1\}$ is a same-hospital indicator,
equal to $1$ if population members $i$ and $j$ were in the same hospital during the epidemic and $0$ otherwise.
The example model is equivalent to a two-parameter exponential family with natural parameters $\theta_1 \in \mR$ and $\theta_2 \in \mR$ and sufficient statistics $s_1(\by) = \sum_{i<j}^N s_{i,j,1}(y_{i,j})$ and $s_2(\by) = \sum_{i<j}^N s_{i,j,2}(y_{i,j})$.
If $\theta_2=0$,
the model of \citet{GrWeHu11} reduces to the model of \citet{BrNe02} and \citet{GrWeHu10}.

\section{Shortcomings of parametric population models}
\label{sec:shortcomings}

While the network-based SEIR models reviewed in Section \ref{likelihood.function} are more flexible than classic and lattice-based SIR and SEIR models,
such parametric population models have shortcomings.
Chief among them is the fact that the induced degree distributions are short-tailed.
Here,
the degrees of population members are the numbers of contacts of population members.

For example,
the model of \citet{BrNe02} and \citet{GrWeHu10} assumes that contacts are independent Bernoulli$(\mu)$ ($\mu \in (0, 1)$) random variables.
As a consequence,
the degrees of population members are Binomial$(N-1,\, \mu)$ and approximately Poisson$(N\, \mu)$ distributed provided $N$ is large, $\mu$ is small, and $N\, \mu$ tends to a constant,
as one would expect in sparse population contact networks where the expected degrees of population members are bounded above by a finite constant and hence $\mu$ is a constant multiple of $1 / N$.
The model of \citet{GrWeHu11} allows degree distributions to be longer-tailed than the model of \citet{BrNe02} and \citet{GrWeHu10}---depending on available covariates---but the induced degree distribution may nonetheless be shorter-tailed than the degree distributions of real-world contact networks.
The degree distributions of real-world contact networks are believed to be long-tailed \citep[e.g.,][]{La94,AlBa02,JoHa04,JoHa03,HaJo04}:
e.g.,
in networks of sexual contacts arising in the study of HIV and other sexually transmitted diseases,
some population members tend to have many more sexual contacts than most other population members.
The population members in the upper tail of the degree distribution represent an important public health concern,
because population members with many contacts can infect many others and are hence potential superspreaders.

Last,
but not least,
it is worth mentioning that scale-free networks with power law degree distributions \citep{BaAl99,AlBa02} are known to induce long-tailed degree distributions.
However,
aside from the fact that the construction of scale-free networks is incomplete and ambiguous \citep*[][]{Boetal01},
those one-parameter models are not flexible models of degree distributions, 
and proper statistical procedures do not lend much support to informal claims that the degree distributions of many real-world contact networks are scale-free:
see,
e.g.,
the work of \citet{JoHa04,JoHa03,HaJo04} on the degree distributions of sexual contact networks arising in the study of HIV, 
and the discussion of \citet*{scalefree}.

Therefore,
more flexible population models are needed to accommodate both short- and long-tailed degree distributions.

\section{Semiparametric population model}
\label{population.model}

We introduce a semiparametric population model to accommodate both short- and long-tailed degree distributions and detect potential superspreaders.
The population model is semiparametric in that the prior of the epidemiological parameter $\bta$ is parametric,
while the prior of the network parameter $\btheta$ is nonparametric.

Let $s_1(\by), \dots, s_N(\by)$ be the degrees of population members $1, \dots, N$,
where the degree of population member $i$ is defined as $s_i(\by) = \sum_{j=1:\, j \neq i}^N\, y_{i,j}$ ($i = 1, \dots, N$).
A simple model of the sequence of degrees $s_1(\by), \dots, s_N(\by)$ is given by the exponential family of distributions with probability mass functions of the form
\bea
\label{degree_model}
p_{\btheta}(\by)
 &=& \exp\left(\dis\sum_{i = 1}^N \theta_i\, s_i(\by) - \psi(\btheta)\right),
 & \by \in \mY,
\eea
where the degrees $s_1(\by), \dots, s_N(\by)$ are the sufficient statistics and the weights of the degrees $\theta_1, \dots, \theta_N \in \mR$ are the natural parameters of the exponential family,
and $\psi(\btheta)$ ensures that $\sum_{\by \in \mY}\, p_{\btheta}(\by) = 1$.
The exponential-family form of \eqref{degree_model} can be motivated by its maximum entropy property and other attractive mathematical properties \citep{Su19}.
A convenient property is that the resulting likelihood function factorizes as follows:
\bea
\label{degree_model_rewritten}
L(\btheta;\, \by)
 &\propto& \exp\left(\dis\sum_{i = 1}^N \theta_i\, s_i(\by) - \psi(\btheta)\right)
 &\propto& \dis\prod_{i < j}^N \exp\left(\lambda_{i,j}(\btheta)\, y_{i,j} - \psi_{i,j}(\btheta)\right),
\eea
where 
\bea
\psi(\btheta)
\= \dsum_{i<j}^N \psi_{i,j}(\btheta),
\eea
with $\psi_{i,j}(\btheta)$ and $\lambda_{i,j}(\btheta)$ defined by
\bea
\psi_{i,j}(\btheta) 
&=& \log\left(1 + \exp\left(\lambda_{i,j}(\btheta)\right)\right)
\eea
and
\bea
\label{log.odds}
\lambda_{i,j}(\btheta) 
&=& \theta_i + \theta_j,
\eea
respectively.

To interpret the natural parameters $\theta_1, \dots, \theta_N$ of the exponential family,
observe that the model is equivalent to assuming that the contacts $Y_{i,j}$ are independent Bernoulli$(\mu_{i,j})$ ($\mu_{i,j} \in (0, 1)$) random variables with $\mbox{logit}(\mu_{i,j}) = \theta_i + \theta_j \in \mR$.
Thus,
the log odds of the probability of a contact between population members $i$ and $j$ is additive in the propensities of $i$ and $j$ to be in contact with others.
It is worth noting that the resulting model can be viewed as an adaptation of the classic $p_1$-model \citep{HpLs81} to undirected random graphs and is known as the $\beta$-model \citep*{DiChSl11,RiPeFi13,ChKaLe19}.

To cluster population members based on degrees and detect potential superspreaders,
we assume that the degree parameters $\theta_1, \dots, \theta_N$ are generated by a Dirichlet process prior \citep{Fe73,Teh2007},
that is,
\bea
\theta_1 &\sim& G\s
 \\
 \theta_{i} \mid \theta_1, \dots, \theta_{i-1} 
 &\sim&
 \dis\frac{1}{\alpha + i - 1} \dis\left(\alpha\, G + \dsum_{h=1}^{i-1} \delta_{\theta_h}\right),\; i = 2, 3, \dots,
\eea
where $\alpha > 0$ denotes the concentration parameter and $G$ denotes the base distribution of the Dirichlet process prior,
while $\delta_{\theta_h}$ denotes a point mass at $\theta_h$. 
A convenient choice of the base distribution $G$ is $N(\mu,\, \sigma^{2})$ ($\mu \in \mR$, $\sigma^2 \in \mR_+$).

Draws from a Dirichlet process prior can be generated by generating 
\bi
\item the first draw from $G$;
\item the $i$-th draw with a probability proportional to $\alpha$ from $G$,
and otherwise drawing one of the existing draws $\theta_1, \dots, \theta_{i-1}$ at random.
\ei

\subsection{Detecting potential superspreaders}

\alert{The described construction of Dirichlet process priors reveals that sampling the degree parameters $\theta_1, \dots, \theta_N$ from a Dirichlet process prior implies that some of the degree parameters are resampled.
As a consequence,
some population members share the same degree parameters.
Thus,
Dirichlet process priors induce a partition of the population into subpopulations,
where subpopulations share the same degree parameters.
In applications to real-world data,
we can make probabilistic statements about which population members belong to subpopulations with high degree parameters based on the posterior distribution.
A short demonstration is provided by the application to the partially observed MERS epidemic in South Korea presented in Section 8.4.}

\subsection{Short- and long-tailed degree distributions}

\begin{figure}[t]
\caption{\label{dpp.plot}
Three sets of draws $\theta_1, \dots, \theta_{1000}$ from the Dirichlet process prior with concentration parameter $\alpha=5$ and base distribution $N(-5,\, 25)$.
Left: Distribution of degree parameters $\theta_1, \dots, \theta_{1000}$.
Right: Distribution of expected degrees $\mu_1(\btheta), \dots, \mu_{1000}(\btheta)$.
Each row corresponds to one set of draws from the Dirichlet process prior.}\s
\begin{center}
\includegraphics[scale=.7]{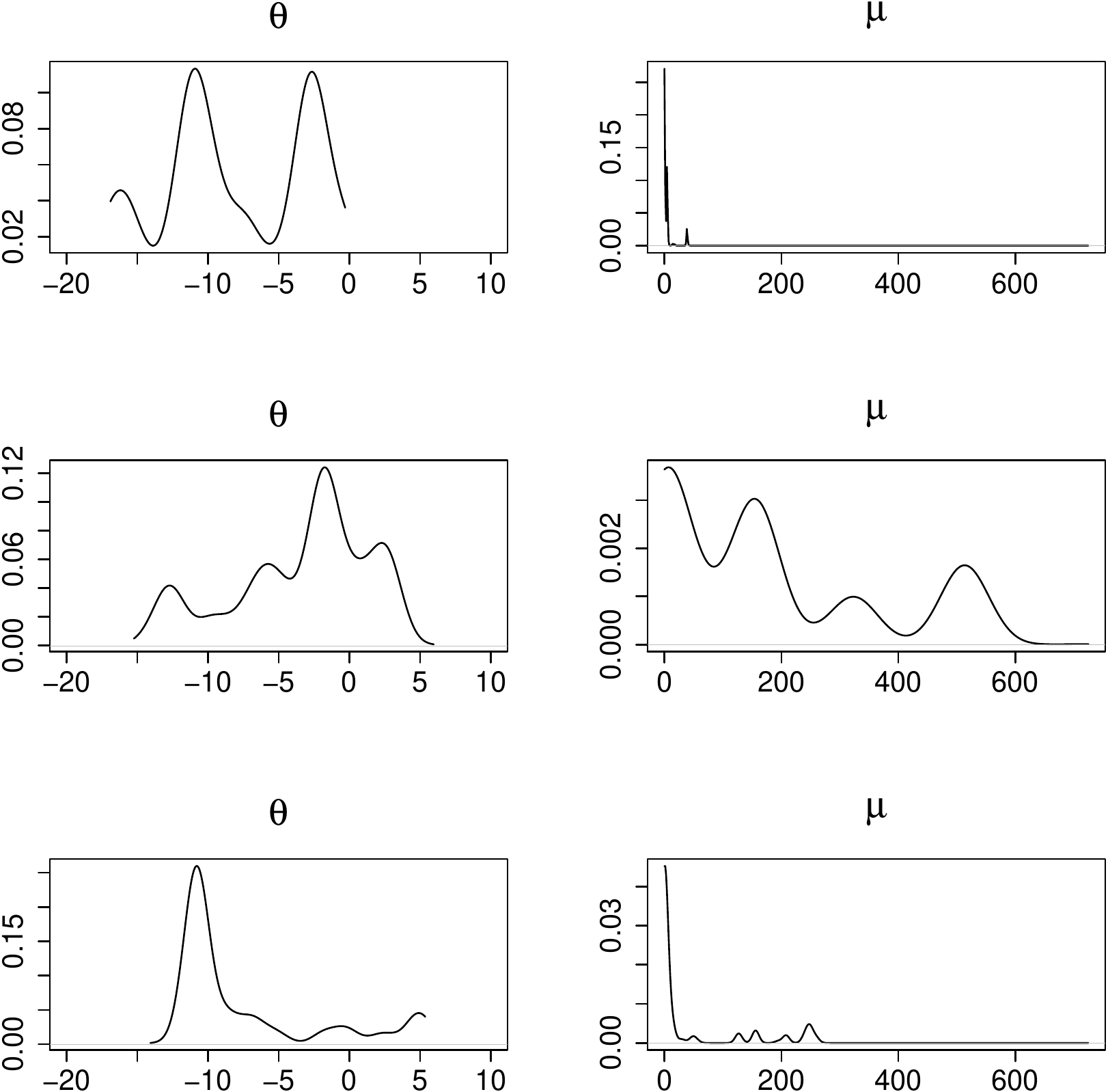}
\end{center}
\end{figure}

\alert{In addition to detecting potential superspreaders,
the model can accommodate short- and long-tailed degree distributions.
To demonstrate that Dirichlet process priors can accommodate short- and long-tailed degree distributions,
we consider a population of size $N = $ 1,000 and generate three sets of degree parameters $\theta_1, \dots, \theta_{1000}$\, from the Dirichlet process prior with 
concentration parameter $\alpha=5$ and base distribution $N(-5,\, 25)$.
Figure 2 shows kernel density plots of the three sets of degree parameters $\theta_1, \dots, \theta_{1000}$ along with the expected degrees of population members.
The expected degree of population member $i$ is defined as
\bea
\mu_i(\btheta)
&=& \mbE_{\btheta}\left(\dsum_{j=1:\, j \neq i}^{1000} Y_{i,j}\right)
&=& \dsum_{j=1:\, j \neq i}^{1000} \dfrac{1}{1 + \exp(-\theta_i-\theta_j)},
& i = 1, \dots, N.
\eea
\hide{
[1] "theta:"
    Min.  1st Qu.   Median     Mean  3rd Qu.     Max. 
-16.9059 -11.1971  -9.8041  -7.9952  -2.7154  -0.2946 
[1] "mean degrees:"
    Min.  1st Qu.   Median     Mean  3rd Qu.     Max. 
 0.00000  0.00096  0.00386  3.98418  4.47277 38.69275 
      75
 4.472773  4.472773  4.472773  4.472773 38.692754 
[1] "theta:"
    Min.  1st Qu.   Median     Mean  3rd Qu.     Max. 
-15.2407  -6.4523  -1.7174  -3.3954   0.5948   5.9689 
[1] "mean degrees:"
    Min.  1st Qu.   Median     Mean  3rd Qu.     Max. 
  0.0007   4.3531 157.5882 175.3936 324.2115 722.5930 
     75
324.2115 324.2115 513.3448 513.3448 513.3448 
[1] "theta:"
   Min. 1st Qu.  Median    Mean 3rd Qu.    Max. 
-14.079 -10.760 -10.760  -7.245  -5.686   5.369 
[1] "mean degrees:"
     Min.   1st Qu.    Median      Mean   3rd Qu.      Max. 
  0.01154   0.31797   0.31797  47.77933  34.57946 259.14991 
      75
 34.57946 127.52472 155.90042 240.41126 248.45654 
}
Figure 2 demonstrates that the distribution of the expected degrees $\mu_1(\btheta), \dots, \mu_{1000}(\btheta)$ can be short- or long-tailed,
depending on the degree parameters $\btheta = (\theta_1, \dots, \theta_{1000})$.
The first set of degree parameters $\theta_1, \dots, \theta_{1000}$ generated from the Dirichlet process prior consists of three subsets of degree parameters,
all of them negative.
The resulting distribution of the expected degrees resembles a steep mountain with a high peak in a neighborhood of $0$ and a short upper tail.
In fact,
90\% of all population members have expected degrees of less than 5,
and the highest expected degree is less than 39,
which is much lower than the highest possible degree of 999 in a population of size 1,000.
The second set of generated degree parameters $\theta_1, \dots, \theta_{1000}$ consists of many negative degree parameters and some positive degree parameters between $0$ and $5$.
Since the log odds of the probability of a contact between two population members $i$ and $j$ is $\theta_i + \theta_j$,
population members with positive degree parameters can have high to very high expected degrees.
Figure 2 shows that the population consists of at least three subpopulations:
population members with expected degrees of less than 100;
population members with expected degrees between 100 and 200;
and population members with expected degree of more than 300.
The highest expected degree is about 722.
The resulting distribution of the expected degrees is both multimodal and long-tailed.
The third set of generated degree parameters $\theta_1, \dots, \theta_{1000}$ resembles the second set of generated degree parameters,
in that the distribution of the expected degrees is multimodal and long-tailed.
That said,
the third set of draws is less extreme than the second set of draws:
e.g.,
the highest expected degree is about 259 rather than 722.

The examples presented above demonstrate that Dirichlet process priors with Gaussian base distributions can accomodate both short- and long-tailed degree distributions,
despite the fact that Gaussians are symmetric and unimodal distributions with light tails.
Other, non-Gaussian base distributions can be chosen.
As a consequence,
the proposed semiparametric population model is flexible and can accomodate a wide range of degree distributions with countless forms and shapes,
including short- and long-tailed degree distributions.
}

\section{Incomplete data}
\label{importance}

\alert{In practice,
complete observations of population contact networks and epidemics are rare.
While population-level data (e.g., counts of the total number of infected, recovered or deceased population members) may be disseminated by public health authorities and can be collected by scraping websites and other channels of communication,
collecting individual-level data (e.g., the contacts, exposure, infectious, and removal times of infected population members) requires substantial investments in terms of time and resources,
making it all but impossible to collect all relevant data.
We first discuss possible reasons for incomplete data in Section 5.1
and then stress the importance of collecting network data in Section 5.2.
}

\vspace{-.25cm}

\subsection{Possible reasons for incomplete data}
\label{data.issue}

\alert{There are many reasons for the fact that available data are,
more often than not,
incomplete.
Some of the possible reasons are:
\bi
\item Epidemics are rare events that occur at random times and in random places,
and when such rare events do occur,
public health officials and scientists may not be well-prepared to collect relevant data without advance notice.
\item Ethical and legal considerations can make the collection of data on individual population members challenging,
if not impossible:
e.g.,
if there was universal cell phone coverage and all population members carried cell phones at all times,
collecting data on contacts among population members would be straightforward by monitoring the locations of cell phones.
However,
collecting such data would violate laws that protect the privacy of population members.
\item Epidemics are not limited to urban areas with excellent infrastructure and ready access to public resources,
but may occur in remote corners of the planet:
e.g.,
the most recent outbreaks of Ebola started in remote areas of Africa.
Worse,
some areas with outbreaks may be war-torn.
As a result,
researchers may not be able to collect data by visiting areas with outbreaks without exposing themselves and others to unacceptable risks.
\item In addition, 
there are more mundane reasons for incomplete data,
such as
\bi
\item[---] {\em design-based mechanisms:} 
e.g.,
motivated by financial constraints,
researchers may sample population members, 
which implies that a sampling design determines which data are collected;
\item[---] {\em out-of-design mechanisms:} 
e.g.,
population members refuse to share data when the data are considered sensitive.
\ei
\ei
}

\vspace{-.35cm}

\subsection{Importance of collecting network data}
\label{sample.networks}

\alert{Data can be incomplete because transmissions, exposure, infectious, or removal times of infected population members are unobserved,
but many available data sets share one fundamental weakness:
There are no data on contacts among population members.
The lack of network data is all the more striking,
because collecting network data would help reduce the posterior uncertainty about
\bi
\item[(a)] the population contact network $\bY$,
which imposes hard constraints on how an infectious disease can spread,
because 
$Y_{i,j}=0$ (no contact) implies  $T_{i,j} = T_{j,i} = 0$ (no transmission) with probability $1$;
\item[(b)] the parameter $\btheta$ of the population model that generated the population contact network $\bY$;
\item[(c)] possible sources of infections.
\ei
Advantages (a) and (b) may not be too surprising,
but advantage (c) may be less obvious.
To demonstrate that sampling contacts can reduce the uncertainty about possible sources of infections,
it is instructive to inspect the full conditional probability of the event that a population member $i$ infected a population member $j$,
assuming that both of them were infected during the course of the epidemic.
If $\varphi(\mbox{$i$ infected $j$})$ denotes the prior probability of the event that population member $i$ infected population member $j$,
then the conditional probability of the event that population member $i$ infected population member $j$,
given everything else, 
takes the form
\bea
\nonumber
\mbP(\mbox{$i$ infected $j$} \mid \bE,\, \bI,\, \bR,\, \bY=\by)
&=& \dfrac{y_{i,j}\; 1_{I_i < E_j < R_i}\; \varphi(\mbox{$i$ infected $j$})}{\sum_{h=1:\, h \neq j}^M\, y_{h,j}\; 1_{I_h < E_j < R_h}\; \varphi(\mbox{$h$ infected $j$})}.
\eea
If $y_{1,j}$, $\dots$, $y_{M,j}$ are observed,
then $y_{1,j}$, $\dots$, $y_{M,j}$ are fixed and impose hard constraints on who could have infected $j$:
If $i$ was not in contact with $j$ (that is, $y_{i,j}=0$),
$i$ could not have infected $j$.
By contrast,
if $y_{1,j}$, $\dots$, $y_{M,j}$ are unobserved,
then $y_{1,j}$, $\dots$, $y_{M,j}$ are not fixed and need to be inferred,
increasing the uncertainty about the possible sources of infection.
In other words,
observed contacts help narrow down the possible sources of infections and, in so doing, help reduce the uncertainty about possible sources of infections.
}

We describe two sampling designs for generating samples of contacts along with epidemiological data:
ego-centric sampling and link-tracing.
Some background on ego-centric sampling and link-tracing of contacts (but not epidemiological data) can be found in 
\citet{ThFr00}, 
\citet{HaGi09},
and \citet{KrMo17}.
\alert{In the literature on network sampling \citep{Fr88b,ThFr00,GiHa06,HaGi09},
popular forms of link-tracing are snowball sampling \citep{Go61} and respondent-driven sampling \citep*{He97,SaHe04,GiHa10,Gi11,kurant.et.al:jsac:2011}.
Some of them do not generate probability samples in the strict sense of the word, 
but generate approximate probability samples \citep[e.g.,][]{Gi11}.
}A recent review of these and other network sampling designs can be found in \citet[][]{ScKrBu17}.
We adapt these ideas to sampling contacts along with epidemiological data.

An ego-centric sample of contacts and epidemiological data can be generated as follows:
\bi
\item[(a)] Generate a probability sample of population members.
\item[(b)] For each sampled population member,
collect data on the contacts of the population member and,
should the population member be infected,
data on possible sources of infection along with the exposure, infectious, and removal times of the population member.
\ei
A probability sample of population members can be generated by any sampling design for sampling from finite populations \citep[e.g.,][]{Th12}.

An interesting extension of ego-centric sampling is link-tracing.
Link-tracing exploits the observed contacts of sampled population members to include additional population members into the sample.
A $k$-wave link-tracing sample of contacts and epidemiological data can be generated as follows:
\bi
\item[(1)] Wave $l = 0$: Generate an ego-centric sample.
\item[(2)] Wave $l = 1, \dots, k$:
\bi
\item[(a)] Add the population members who are connected to the population members of wave $l - 1$ to the sample.
\item[(b)] For each added population member,
collect all relevant data.
\ei
\ei
Ego-centric sampling can be considered to be a special case of $k$-wave link-tracing with $k = 0$.
We use ego-entric sampling in the simulation study in Section \ref{simulation.study}.

\section{Bayesian inference}
\label{bay}

\alert{We discuss Bayesian inference for the parameters $\bta$ and $\btheta$ of the population model based on incomplete data.
Since interest centers on the population model,
it is natural to ask:
Under which conditions is the process that determines which data are observed ignorable for the purpose of Bayesian inference for the parameters $\bta$ and $\btheta$ of the population model?
To answer the question,
we start from first principles.
We first separate the complete-data generating process from the incomplete-data generating process:
\bi
\item The {\em complete-data generating process} is the process that generates the complete data,
that is,
the process that generates a realization $(\bx, \by)$ of $(\bX, \bY)$.
\item The {\em incomplete-data generating process} is the process that determines which subset of the complete data $(\bx, \by)$ is observed.
\ei
A failure to separate these processes can lead to misleading conclusions,
as pointed out by \citet{Ru76}, \citet{DaDi77}, \citet*{KoRoPa09}, \citet{HaGi09,GiHa17}, \citet{crane2018probabilistic}, and \citet{ScKrBu17}.

We therefore proceed as follows.
We first separate the complete-data generating process from the incomplete-data generating process in Section 6.1
and then discuss Bayesian inference based on incomplete data in Section 6.2.
We then discuss Bayesian computing in Sections 6.3, 6.4, and 6.5.
}

\subsection{Complete- and incomplete-data generating process}
\label{likelihood.ignorable}

\alert{To prepare the ground for principled Bayesian inference,
we separate the complete-data generating process from the incomplete-data generating process,
adapting the generic ideas of \citet{Ru76} to stochastic models of epidemics.}

To do so,
let $\bA = \{\bA_E, \bA_I, \bA_R, \bA_T, \bA_Y\}$ be indicators of which data are observed,
where $\bA_E = \{A_{E,i}\}_{i=1}^N \in \{0, 1\}^N$,
$\bA_I = \{A_{I,i}\}_{i=1}^N \in \{0, 1\}^N$,
$\bA_R = \{A_{R,i}\}_{i=1}^N \in \{0, 1\}^N$,\linebreak
$\bA_T = \{A_{T,i,j}\}_{i \neq j}^M  \in \{0, 1\}^{M^2-M}$, 
and $\bA_Y = \{A_{Y,i,j}\}_{i<j}^N \in \{0, 1\}^{\binom{N}{2}}$ indicate whether the values of 
$\{E_i\}_{i=1}^N$, $\{I_i\}_{i=1}^N$, $\{R_i\}_{i=1}^N$, $\{T_{i,j}\}_{i \neq j}^M$, and $\{Y_{i,j}\}_{i<j}^N$ are observed,
respectively.
The sequence of indicators $\bA$ is considered to be a random variable,
with a distribution parameterized by $\bvartheta \in \Omega_{\bvartheta} \subseteq \mR^q$ ($q \geq 1$):
e.g.,
$\bvartheta \in [0, 1]^N$ may be a vector of sample inclusion probabilities,
where $\vartheta_i \in [0, 1]$ is the probability that population member $i \in \{1, \dots, N\}$ is sampled and data on the contacts of population member $i$ are collected.
The parameter $\bvartheta$ may be known or unknown.
We focus on the more challenging case where $\bvartheta$ is unknown.
The observed and unobserved subset of the complete data $(\bx,\, \by)$ are denoted by $\bx_{\mbox{\tiny obs}}$ and $\bx_{\mbox{\tiny mis}}$ and $\by_{\mbox{\tiny obs}}$ and $\by_{\mbox{\tiny mis}}$,
respectively,
where $\bx = (\bx_{\mbox{\tiny obs}},\, \bx_{\mbox{\tiny mis}})$ and $\by = (\by_{\mbox{\tiny obs}},\, \by_{\mbox{\tiny mis}})$.

In Bayesian fashion,
we build a joint probability model for all knowns and unknowns and condition on all knowns.
Let
\bea
\label{joint}
p(\baa,\, \bx,\, \by,\, \bvartheta,\, \bta,\, \btheta) 
\= p(\baa,\, \bx,\, \by \mid \bvartheta,\, \bta,\, \btheta)\; p(\bvartheta,\, \bta,\, \btheta)
\eea
be the joint probability density of $\baa$, $\bx$, $\by$, $\bvartheta$, $\bta$, $\btheta$,
where
\bea
\label{joint.prior}
p(\bvartheta,\, \bta,\, \btheta) 
\= p(\bvartheta \mid \bta,\, \btheta)\; p(\bta,\, \btheta)
\eea
is the prior probability density of $\bvartheta$, $\bta$, $\btheta$ and
\bea
\label{joint.pdf}
p(\baa,\, \bx,\, \by \mid \bvartheta,\, \bta,\, \btheta) 
\= p(\baa \mid \bx,\, \by,\, \bvartheta)\, p(\bx \mid \by,\, \bta)\, p(\by \mid \btheta)
\eea
is the conditional probability density of $\baa$, $\bx$, $\by$ given $\bvartheta$, $\bta$, $\btheta$;
\hide{
It is natural to assume that
$p(\baa \mid \bx, \by, \bvartheta, \bta, \btheta) = p(\baa \mid \bx, \by, \bvartheta)$ does not depend on $\bta$ and $\btheta$,
$p(\bx \mid \by, \bvartheta, \bta, \btheta) = p(\bx \mid \by, \bta)$ does not depend on $\bvartheta$ and $\btheta$, 
and $p(\by \mid \bvartheta, \bta, \btheta) = p(\by \mid \btheta)$ does not depend on $\bvartheta$ and $\bta$;
}
note that $p(\by \mid \btheta) \equiv p_{\btheta}(\by)$.
It is worth noting that all of these probability densities are with respect to a suitable dominating measure,
but we do not wish to delve into measure-theoretic details,
which would distract from the main ideas.

\alert{To determine when the incomplete-data generating process is ignorable for the purpose of Bayesian inference for the parameters $\bta$ and $\btheta$ of the population model,
we introduce the following definition of a likelihood-ignorable incomplete-data generating process.}\s

{\bf Definition: likelihood-ignorable incomplete-data generating process.}
{\em 
Assume that the parameters $\bvartheta$, $\bta$, $\btheta$ are variation-independent in the sense that the parameter space of $(\bvartheta,\, \bta,\, \btheta)$ is given by a product space of the form $\Omega_{\bvartheta} \times \Omega_{\bta} \times \Omega_{\btheta}$ and that the parameters of the population model $\bta$ and $\btheta$ and the parameter of the incomplete-data generating process $\bvartheta$ are independent under the prior,
\bea
\label{joint.prior}
p(\bvartheta \mid \bta,\, \btheta) 
\= p(\bvartheta) & \mbox{for all} & (\bvartheta,\, \bta,\, \btheta) \in \Omega_{\bvartheta} \times \Omega_{\bta} \times \Omega_{\btheta}.
\eea
If the probability of observing data does not depend on the values of the unobserved data,
\bea
\label{conditional.u}
p(\baa \mid \bx,\, \by,\, \bvartheta)
\;=\; p(\baa \mid \bx_{\mbox{\tiny obs}},\, \by_{\mbox{\tiny obs}},\, \bvartheta) & \mbox{for all} & \baa,\; \bx,\; \by \mbox{ and all } \bvartheta \in \Omega_{\bvartheta},
\eea
then the incomplete-data generating process is called likelihood-ignorable, 
and otherwise non-ignorable.
}

\vspace{-.35cm}

We provide two examples of likelihood-ignorable and non-ignorable incomplete-data generating processes.\s

\alert{{\bf Example: likelihood-ignorable.}}
\alert{{\em All infected population members visit hospitals,
which record data on contacts, transmissions, exposure, infectious, and removal times of infected population members.
To reduce the posterior uncertainty about the population contact network and its generating mechanism,
investigators generate a probability sample of non-infected population members from the subpopulation of all non-infected population members and collect data on the contacts of sampled population members,
rather than limiting the collection of data to infected population members visiting hospitals.}
}\s

\alert{A possible sampling design for generating samples of contacts and epidemiological data is link-tracing,
as described in Section \ref{sample.networks}.
Link-tracing exploits observed contacts of population members to increase the size of the sample,
which implies that the sample inclusion probabilities depend on observed contacts but do not depend on unobserved contacts:
\bea
p(\baa \mid \bx,\, \by,\, \bvartheta)
\;=\; p(\baa \mid \bx_{\mbox{\tiny obs}},\, \by_{\mbox{\tiny obs}},\, \bvartheta) & \mbox{for all} & \baa,\; \bx,\; \by \mbox{ and all } \bvartheta \in \Omega_{\bvartheta}.
\eea
Therefore,
link-tracing sampling designs for generating samples of contacts and epidemiological data are likelihood-ignorable,
as are ego-centric sampling designs.}
\vspace{.25cm}

{\bf Example: non-likelihood-ignorable.}
{\em Suppose that there exists a constant $\delta > 0$ such that infected population members $i$ with mild symptoms and fast recovery ($R_i - I_i \leq \delta$) do not visit hospitals,
whereas infected population members $i$ with severe symptoms and slow recovery ($R_i - I_i > \delta$) do visit hospitals.
Hospitals collect data on infected population members who visit them,
but no data are collected on other population members.\s
}

Since the incomplete-data generating process excludes all infected population members with mild symptoms and fast recovery,
it cannot be ignored.
Statistical analyses ignoring it may give rise to misleading conclusions about the rate of infection $\beta$ and other parameters of the population model,
and may generate misleading predictions of future epidemics in the population of interest and similar populations.

\subsection{Bayesian inference based on incomplete data}
\label{bayesian.inference}

\alert{Separating the complete-data generating process from the incomplete-data generating process paves the way for principled Bayesian inference based on incomplete data.

The following result shows that,
as long as the incomplete-data generating process is likelihood-ignorable,
Bayesian inference for the parameters $\bta$ and $\btheta$ of the population model 
can be based on the marginal posterior $p(\bta,\, \btheta \mid \baa,\, \bx_{\mbox{\tiny obs}},\, \by_{\mbox{\tiny obs}})$.
In other words,
the incomplete-data generating process can be ignored for the purpose of Bayesian inference for the parameters $\bta$ and $\btheta$ of the population model,
because the marginal posterior $p(\bta,\, \btheta \mid \baa,\, \bx_{\mbox{\tiny obs}},\, \by_{\mbox{\tiny obs}})$ can be computed without requiring knowledge about the incomplete-data generating process.
}

\vspace{-.85cm}

\alert{
\proposition\label{theorem.2}
{\em If the incomplete-data generating process is likelihood-ignorable and the prior $p(\bta,\, \btheta)$ is proper,
the parameter $\bvartheta$ of the incomplete-data generating process and the parameters $\bta$ and $\btheta$ of the population model are independent under the posterior,
\bea
\nonumber
\label{result1}
p(\bvartheta,\, \bta,\, \btheta \mid \baa,\, \bx_{\mbox{\tiny obs}},\, \by_{\mbox{\tiny obs}})
&\propto& p(\bvartheta \mid \baa,\, \bx_{\mbox{\tiny obs}},\, \by_{\mbox{\tiny obs}})\; p(\bta,\, \btheta \mid \baa,\, \bx_{\mbox{\tiny obs}},\, \by_{\mbox{\tiny obs}}),
\eea
and Bayesian inference for the parameters $\bta$ and $\btheta$ of the population model can be based on the marginal posterior $p(\bta,\, \btheta \mid \baa,\, \bx_{\mbox{\tiny obs}},\, \by_{\mbox{\tiny obs}})$,
\bea
\label{main.result}
p(\bta,\, \btheta \mid \baa,\, \bx_{\mbox{\tiny obs}},\, \by_{\mbox{\tiny obs}})
= \dfrac{\sum_{\by_{\mbox{\tiny mis}}} \int p(\bx \mid \by,\, \bta)\; p(\by \mid \btheta)\; p(\bta,\, \btheta) \dd \bx_{\mbox{\tiny mis}}}{\sum_{\by_{\mbox{\tiny mis}}} \int\int\int p(\bx \mid \by,\, \bta)\, p(\by \mid \btheta)\, p(\bta,\, \btheta) \dd \bx_{\mbox{\tiny mis}} \dd \bta \dd \btheta},
\eea
which can be computed without requiring knowledge about the incomplete-data generating process.}
}

\s

\alert{It is worth noting that the unobserved contacts of non-infected population members can be summed out,
both in the numerator and the denominator of the ratio in \eqref{main.result},
because the contacts of population members are independent conditional on $\btheta$ by construction of the semiparametric population model described in Section 4.
In other words,
there is no need to use computational methods (e.g., Markov chain Monte Carlo methods) to generate model-based imputations of unobserved contacts of non-infected population members,
which saves computing time.}

A proof of Proposition 1 can be found in the supplement.
The case of \citet{BrNe02} and \citet{GrWeHu10,GrWeHu11},
who considered Bayesian inference from observed infectious and removal times, 
is a special case of the incomplete-data framework considered here, 
with $\bx_{\mbox{\tiny obs}} = \{\bI, \bR\}$ and $\by_{\mbox{\tiny obs}} = \{\}$.
In general,
when $\bx$ or $\by$ are partially observed,
Bayesian inference can be based on the marginal posterior of $\bta$ and $\btheta$ given the observed data,
provided the incomplete-data generating process is likelihood-ignorable.

\subsection{Truncated Dirichlet process priors}
\label{truncation}

\alert{To facilitate Markov chain Monte Carlo sampling from the posterior distribution,\linebreak
we approximate Dirichlet process priors by truncated Dirichlet process priors along the lines of \citet{IsJa01}.

The truncation of Dirichlet process priors takes advantage of the stick-breaking construction of Dirichlet process priors \citep{Se94,IsJa01,Teh2007}.
A stick-breaking construction of a Dirichlet process prior with base distribution $G$ and concentration parameter $\alpha$ proceeds as follows.
First,
we sample parameters from the base distribution $G$: 
\bea
\nonumber
\gamma_k &\msim\limits^{\tiny \mbox{iid}}& G,
& k = 1, 2, \dots
\eea
We then construct mixing proportions $\pi_1$, $\pi_2$, $\dots$ by first sampling
\bea
\nonumber
  V_k \mid \alpha &\msim\limits^{\tiny \mbox{iid}}& \mbox{Beta}(1, \alpha),
& k = 1, 2, \dots
\eea
and then setting
\bea
\nonumber
 \pi_1 & = & V_1\s
 \\
 \pi_k & = & V_k\, \dis\prod_{j = 1}^{k - 1} (1 - V_j),\; k = 2, 3, \dots
\eea
Last,
but not least,
we construct an infinite mixture distribution with mixing proportions $\pi_1, \pi_2, \dots$ and point masses $\delta_{\gamma_1}$, $\delta_{\gamma_2}$, $\dots$ as follows:
\bea
\nonumber
\mathscr{P}_\infty &=& \dsum_{k=1}^\infty\, \pi_k\; \delta_{\gamma_k}.
\eea
The distribution $\mathscr{P}_\infty$ is then a draw from the Dirichlet process prior with concentration parameter $\alpha$ and base distribution $G$. 

A Dirichlet process prior can be truncated by choosing a positive integer $K$ and setting $V_K = 1$,
which implies that $\pi_{K+1} = 0$, $\pi_{K+2} = 0$, $\dots$
A finite mixture distribution with mixing proportions $\pi_1, \pi_2, \dots, \pi_K$ and point masses $\delta_{\gamma_1}$, $\delta_{\gamma_2}$, $\dots$, $\delta_{\gamma_K}$ can then be constructed as follows:
\bea
\nonumber
\mathscr{P}_K &=& \dsum_{k=1}^K\, \pi_k\; \delta_{\gamma_k}.
\eea
The distribution $\mathscr{P}_K$ can be regarded as a draw from the Dirichlet process prior with concentration parameter $\alpha$ and base distribution $G$ truncated at $K$.
If $K$ is large,
the truncated Dirichlet process prior is expected to be a good approximation of the Dirichlet process prior.
Some theoretical guidance regarding the choice of $K$ can be found in \citet{IsJa01}.
In practice,
$K$ can be chosen by 
\bi
\item selecting a large positive integer, 
such as $K = N$;
\item exploiting domain knowledge;
\item choosing a value of $K$ that leads to acceptable in- or out-of-sample performance.
\ei
The truncated stick-breaking construction of $\bpi$ implies that $\bpi$ is generalized Dirichlet distributed,
which is conjugate to multinomial sampling \citep{IsJa01} and facilitates Markov chain Monte Carlo sampling from the posterior distribution.

Since the concentration parameter $\alpha$ and the parameters $\mu$ and $\sigma^2$ of the base distribution $N(\mu,\, \sigma^2)$ are unknown,
it is natural to express the uncertainty about $\alpha$, $\mu$, and $\sigma^2$ by assuming that $\alpha$, $\mu$, and $\sigma^2$ have hyperpriors.
We assume that the hyperpriors of the hyperparameters $\alpha$, $\mu$, and $1 / \sigma^{2}$ are Gamma, Gaussian, and Gamma distributions,
respectively,
which are conjugate priors and facilitate Markov chain Monte Carlo sampling from the posterior distribution.

}

\subsection{Bayesian Markov chain Monte Carlo algorithm}
\label{mmcc}

\alert{A Bayesian Markov chain Monte Carlo algorithm for sampling from the posterior distribution is described in the supplement.
We list here the main steps of the algorithm,
without providing details,
and discuss its computing time.
Details can be found in the supplement.

To list the main steps of the algorithm,
let
\bi
\item $Z_{i,k} = 1$ if population member $i$ was assigned degree parameter $\gamma_k$ and $Z_{i,k}= 0$ otherwise ($i = 1, \dots, N$,\; $k = 1, \dots, K$);
\item $\bZ_i = (Z_{i,1}, \dots, Z_{i,K})$ and $\bZ_{-i} = (\bZ_1, \dots, \bZ_{i-1}, \bZ_{i+1}, \dots, \bZ_N)$ ($i = 1, \dots, N$),
and $\bZ = (\bZ_1, \dots, \bZ_N)$;
\item $\bgamma = (\gamma_1, \dots, \gamma_K)$;
\item $\bpi = (\pi_1, \dots, \pi_K)$.
\ei
As a consequence,
the degree parameter $\theta_i$ of population member $i$ can be expressed as 
\bea
\nonumber
\theta_i 
\= \bZ_i^\top\, \bgamma,
&& i = 1, \dots, N.
\eea
A Markov chain Monte Carlo algorithm for sampling from the posterior distribution can then be constructed by combining the following Markov chain Monte Carlo steps by means of cycling or mixing \citep*{Tl94,Li08},
assuming all unknown quantities have been initialized:
\begin{enumerate}
\item Impute the missing data:
\bi
\item Sample $\bX_{\mbox{\tiny mis}} \mid \bX_{\mbox{\tiny obs}} = \bx_{\mbox{\tiny obs}},\, \bY = \by,\, \bta$.
\item Sample $\bY_{\mbox{\tiny mis}} \mid \bX=\bx,\, \bY_{\mbox{\tiny obs}} = \by_{\mbox{\tiny obs}},\, \bZ=\bz,\, \bgamma,\, \bta$.
\ei
\item Sample the parameters of the population model:
\bi
\item Sample $\bgamma \mid \bY=\by,\, \bZ=\bz$.
\item Sample $\bZ_i \mid \bY=\by,\, \bZ_{-i} = \bz_{-i},\, \bgamma,\, \bpi$ ($i = 1, \dots, N$).
\item Set $\theta_i = \bZ_i^\top\, \bgamma$ ($i = 1, \dots, N$).
\item Sample $\bta \mid \bX = \bx,\, \bY = \by$.
\ei
\item Sample the hyperparameters:
\bi
\item Sample $\alpha \mid \bpi$.
\item Sample $\bpi \mid \bZ=\bz,\, \alpha$.
\item Sample $\mu \mid \sigma^2,\, \bgamma$.
\item Sample $\sigma^{2} \mid \mu,\, \bgamma$.
\ei
\end{enumerate}
\hide{
Steps 1, 2, and 3 sample from the full conditional Gamma, Gaussian, and Gamma distributions of $\alpha$, $\mu$, and $\sigma^{-2}$.
Step 4 amounts to sampling stick-breaking weights $V_1, \dots, V_{K-1}$ from full conditional Beta distributions and setting $V_K = 1$,
then constructing $\bpi$ by using the stick-breaking construction described in Section 6.3.
}
Most of the Markov chain Monte Carlo steps involve Gibbs sampling from full conditional distributions (e.g., Beta, Gamma, and Gaussian distributions),
while the others are Metropolis-Hastings steps.
More details are provided in the supplement.

The computing time of the algorithm is a function of
\bi
\item the number of subpopulations $K$, 
which satisfies $K \leq N$;
\item the number of infected population members $M$, 
which satisfies $M \leq N$ and in large populations satisfies $M \ll N$ (unless a non-negligible fraction of the population is infected);
\item the sampling design and the number of population members $n$ sampled out of the $N$ population members for the purpose of collecting data on contacts along with epidemiological data,
which satisfies $n \leq N$ and in large populations satisfies $n \ll N$ (unless a non-negligible fraction of the population is sampled);
\item the sparsity of the population contact network.
\ei
As a specific example,
consider ego-centric sampling, 
as described in Section 5.2.
An ego-centric sampling design samples $n$ out of the $N$ population members and,
for each sampled population member,
collects data on the contacts of the sampled population member with the $N-1$ other population members,
in addition to data on the transmissions, exposure, infectious, and removal times of infected population members.
As a result,
the computing time of each iteration of the Bayesian Markov chain Monte Carlo algorithm is $O(K\, n\, N)$,
because updates of the $K$ degree parameters $\gamma_1, \dots, \gamma_K$ involve computations of up to $n\, (N-1)$ probabilities of the form 
\bea
\nonumber
\mbP_{\theta_i,\theta_j}(Y_{i,j}=y_{i,j})
&=& \exp((\theta_i+\theta_j)\, y_{i,j} - \log(1 + \exp(\theta_i+\theta_j))),
\eea
where $\theta_i = \bZ_i^\top\, \bgamma$ and $\theta_j = \bZ_j^\top\, \bgamma$.

Having said that,
the computing time of $O(K\, n\, N)$ in the case of ego-centric sampling is based on worst-case scenarios and can be reduced to $O(K\, n)$ in special cases:
e.g.,
when the population contact network is sparse in the sense that many population members have few contacts,
it is possible to reduce the computing time by taking advantage of sparsity.
In fact,
many real-world networks are sparse:
While population members can create up to $N-1$ contacts,
creating physical contacts that enable disease transmission requires geographical proximity and time.
As a consequence,
it is plausible that the expected degrees of many if not all population members are bounded above by a finite constant \citep*{Du92,RoChYu11,KrHaMo11,Lo12,AmChBiLe13,KrKo14,butts:jms:2018}.
In such cases,
the resulting population contact network is sparse, 
and one could reduce the worst-case computing time of $O(K\, n\, N)$ to $O(K\, n)$.
Some ideas of how to exploit sparsity for the purpose of reducing computing time can be found in, e.g., \citet*{RaNiHoYe12} and \citet*{VuHuSc12}.
}

\subsection{Addressing the label-switching problem}

\alert{Markov chain Monte Carlo samples from the posterior distribution may show evidence of label-switching,
that is,
the labeling of subpopulations may have switched from iteration to iteration of the Markov chain Monte Carlo algorithm.
The label-switching problem is rooted in the fact that the likelihood function is invariant to permutations of the labels of subpopulations.
While the prior is not invariant to permutations of the labels of subpopulations,
the prior is dominated by the likelihood function when the data is informative.
As a consequence,
the labels of subpopulations may switch from iteration to iteration of the Markov chain Monte Carlo algorithm.
Label-switching can give rise to misleading conclusions about parameters that depend on the labeling of the subpopulations,
including the indicators $\bZ_1, \dots, \bZ_N$ and the degree parameters $\gamma_1, \dots, \gamma_K$ and $\theta_1 = \bZ_1^\top\, \bgamma, \dots, \theta_N = \bZ_N^\top\, \bgamma$.
The label-switching problem is a well-known problem in the literature concerned with Bayesian Markov chain Monte Carlo estimation of finite mixture models and related models.
We follow the Bayesian decision-theoretic approach of \citet{St00} to undoing the label-switching,
by minimizing the posterior expectation of a well-chosen loss function.
A loss function and relabeling algorithm are stated in \citet{ScHa13} and implemented in {\tt R} package {\tt hergm} \citep{ScLu15}.
We use them in Sections 7 and 8 to undo the label-switching in all Markov chain Monte Carlo samples generated by the Bayesian Markov chain Monte Carlo algorithm.
}

\section{Simulation results}
\label{simulation.study}

\alert{We explore the frequentist properties of Bayesian point estimators and the reduction in statistical error due to collecting network data by using simulations.
We consider a population of size 187 consisting of $K = 3$ subpopulations labeled $1$, $2$, $3$.
The three subpopulations consist of low-, moderate-, and high-degree population members.
We assign population members $i$ to subpopulations $1, 2, 3$ by sampling $\bZ_i \iid \mbox{Multinomial}(1;\, \bpi=(.4,\, .3,\, .3))$ ($i = 1, \dots, N$).
We then generate a population contact network from the population model described in Section 4 with degree parameters $\theta_i = \bZ_i^\top\, \bgamma$ ($i = 1, \dots, N$). 
Conditional on the population contact network,
an epidemic is generated by the stochastic model described in Section 2,
assuming that $I_i - E_i$ and $R_i - I_i$ are independent Gamma$(\eta_{E,1},\, \eta_{E,2})$ and Gamma$(\eta_{I,1},\, \eta_{I,2})$ random variables,
respectively ($i = 1, \dots, M$).
The data-generating values of the parameters are specified in Sections 7.1 and 7.2.
Unless stated otherwise,
we assume that the exposure, infectious, and removal times $\bE$, $\bI$, $\bR$ are observed,
whereas the transmissions $\bT$ are unobserved,
as are the population contact network $\bY$ and the indicators $\bZ$.
}

\subsection{Simulation results quantifying the error of estimation}
\label{sim.recovery}

\vspace{-.5cm}

\begin{table}[h]
\caption{\label{simulation.recovery}
{\normalsize\em Simulation results. 
Number of times 95\% posterior credible intervals include the data-generating values of the parameters $\beta$, $\eta_{E,1}$, $\eta_{E,2}$, $\eta_{I,1}$, $\eta_{I,2}$, $\gamma_1$, $\gamma_2$, and $\gamma_3$ in \%, 
using a Dirichlet process prior truncated at $K = 3$ and $K = 5$.}}\s
\begin{minipage}{6in}
\begin{center}
\begin{tabular}{lcccccccccccccccc}
        Parameter                   & $\beta$               & $\eta_{E,1}$         & $\eta_{E,2}$    &       $\eta_{I,1}$   & $\eta_{I,2}$    & $\gamma_1$            & $\gamma_2$            &       $\gamma_3$ \\[0.5ex]
        \hline
        True value                           &  2                    &       8       &       .25            &       4               &       .25            &       -2                      &       -1                      &       0\\
        \hline
        $K = 3$      &  95.4\%               &  95.4\%       &       93.5\%  &       95.2\%  &       94.3\%  &       83.4\%          &       97.4\%          &       89.4\%\\
        \hline
        $K = 5$     &  93.0\%               & 96.3\%        &       96.3\%  &       96.0\%  &       95.7\%  &       98.4\%          &       100\%           &       96.7\%\\
        \hline
\end{tabular}
\end{center}
\end{minipage}
\end{table}

\hide{






}

\alert{We generate 1,000 population contact networks and epidemics as described above.
The data-generating values of the parameters $\beta$, $\eta_{E,1}$, $\eta_{E,2}$, $\eta_{I,1}$, $\eta_{I,2}$, $\gamma_1$, $\gamma_2$, and $\gamma_3$ are shown in Table \ref{simulation.recovery}.
For each data set,
we truncate the Dirichlet process prior at $K = 3$ and $K = 5$ as described in Section 6.3,
and estimate the data-generating model by the Bayesian Markov chain Monte Carlo algorithm described in Section 6.4.}

\alert{Table \ref{simulation.recovery} sheds light on the frequentist coverage properties of 95\% posterior credible intervals.
The simulation results indicate that the frequentist coverage properties of posterior credible intervals are excellent in the case of the epidemiological parameters $\beta$, $\eta_{E,1}$, $\eta_{E,2}$, $\eta_{I,1}$, and $\eta_{I,2}$,
but less so in the case of the network parameters $\gamma_1$, $\gamma_2$, and $\gamma_3$.
These results underscore the challenge of estimating network parameters without observing network data.
Section \ref{sim.sampling} demonstrates that the statistical error can be reduced by collecting network data.}

\begin{figure}[t]
\caption{\label{mse.plot}{Simulation results: MSE of the posterior median and mean of the parameters $\beta$, $\gamma_1$, $\gamma_2$, and $\gamma_3$ plotted against the sample size $n = 0$,\, $25$,\, $50$,\, $75$,\, $100$,\, $125$,\, $150$,\, $187$.\break
If $n=0$, 
no contacts are observed,
otherwise a subset of contacts is observed.}}
\begin{center}
\includegraphics[scale=.775]{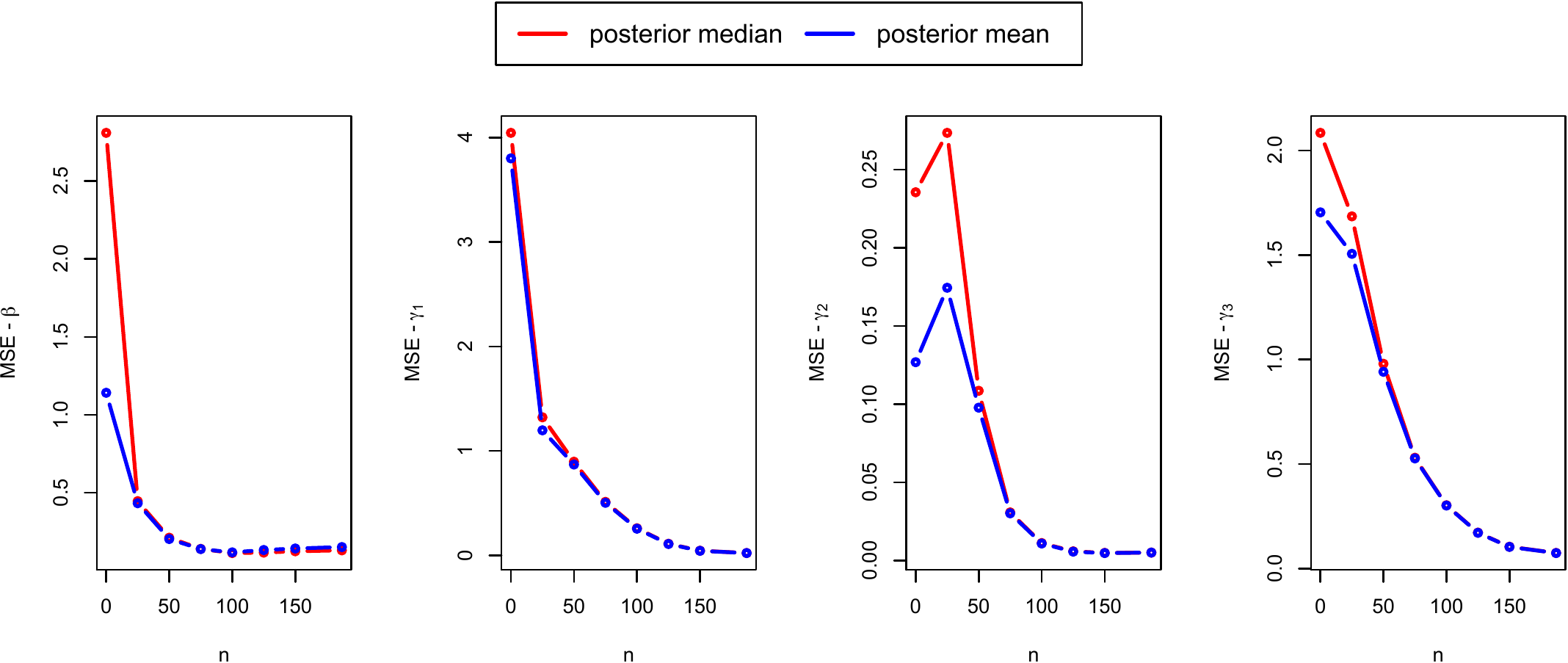}
\end{center}
\end{figure}

\subsection{Simulation results quantifying the effect of network sampling}
\label{sim.sampling}

\alert{To demonstrate that collecting network data can reduce the posterior uncertainty about the parameters of the population model,
we consider a population consisting of $K=3$ subpopulations.
The $K=3$ subpopulations correspond to 
\bi
\item a low-degree subpopulation of size 127 with degree parameter $\gamma_1 = -3.5$;
\item a moderate-degree subpopulation of size 50 with degree parameter $\gamma_2 = -1.5$;
\item a high-degree subpopulation of size 10 with degree parameter $\gamma_3 = .5$.
\ei
We generate 1,000 ego-centric samples of sizes $n = 25$,\, $50$,\, $75$,\, $100$,\, $125$,\, $150$,\, $187$ from the population of size $N = 187$.
We then estimate the population model from each sample of contacts along with observations of the exposure, infectious, and removal times of infected population members.
In addition,
we estimate the population model without observations of contacts,
which corresponds to a sample size of $n=0$,
using observations of the exposure, infectious, and removal times of infected population members.
To assess how much the posterior uncertainty about the parameters of the population model is reduced by sampling contacts,
we use the mean squared error (MSE) of the posterior median and mean of the parameters.
}

\alert{The MSE of the posterior median and mean of the rate of infection $\beta$ and the degree parameters $\gamma_1$, $\gamma_2$, and $\gamma_3$ is plotted in Figure \ref{mse.plot} against the sample size $n = 0$,\, $25$,\, $50$,\, $75$,\, $100$,\, $125$,\, $150$,\, $187$.
By construction of the model,
estimators of the epidemiological parameters $\eta_{E,1}$, $\eta_{E,2}$, $\eta_{I,1}$, and $\eta_{I,2}$ are not expected to be sensitive to $n$---which determines how much information is available about the network parameters---and the MSE of the posterior median and mean of $\eta_{E,1}$, $\eta_{E,2}$, $\eta_{I,1}$, and $\eta_{I,2}$ are indeed not sensitive to $n$ (not shown).
However,
Figure \ref{mse.plot} demonstrates that samples of contacts do reduce the MSE of the posterior median and mean of $\beta$, $\gamma_1$, $\gamma_2$, and $\gamma_3$:
The MSE turns out to be highest when $n = 0$,
and rapidly decreases as $n$ increases.
These observations underscore the importance of collecting data on contacts or functions of contacts, 
such as degrees.}

\section{Partially observed MERS epidemic in South Korea}
\label{application}

\alert{We showcase the statistical framework introduced in Sections 4, 5, and 6 by applying it to the partially observed MERS epidemic in South Korea in 2015 \citep{Ki15}.}

\subsection{Data}
\label{sec:data}

\alert{The MERS outbreak in South Korea was driven by the coronavirus MERS,
which is related to the coronaviruses SARS and COVID-19.
We retrieved the data from the website {\tt http://english.mw.go.kr} of the South Korean Ministry of Health and Welfare on September 22, 2015.
The website has been removed since,
but the data can be obtained from the authors.
The first MERS case in South Korea was confirmed on May 20, 2015 and the last case was confirmed on July 4, 2015.
By September 22, 2015,
186 cases had been confirmed,
of which 144 had recovered while 35 had died,
all of whom are considered to be removed from the population.
7 cases had not removed by September 22, 2015, 
despite the fact that the last case of MERS was confirmed on July 4, 2015.
We assume that the removal times of those 7 cases are unobserved and that the outbreak ceased by September 22, 2015,
because the data base has not been updated by the South Korean Ministry of Health and Welfare since July 2015.
The data consist of the infectious times and removal times of the 186 infected population members, 
with 7 missing removal times.
The exposure times are unobserved and inferred along with the 7 missing removal times.
The MERS data set does not include direct observations of transmissions or contacts,
but there are two sources of indirect observations:
\bi
\item The assessments of doctors of who infected whom.
\item The observed infectious and removal times,
which reveal when infected population members were infectious and,
in so doing, 
help narrow down the possible sources of infections.
\ei
Both of these sources help inform who infected whom and who was in contact with whom,
because an infection implies a contact.
}

\hide{

\alert{Having said that,
these indirect observations have shortcomings:
\bi
\item The assessments of doctors are incomplete,
because the doctors did not specify possible sources of infections for 26 out of the 186 infected population members.
\item To the best of our knowledge,
the assessments of doctors were educated guesses of possible sources of infections.
\item An infection implies a contact,
but a contact may not lead to an infection,
so the number of infections is a lower bound on the number of contacts.
\ei
We use the assessments of doctors concerning who infected whom as a prior for who infected whom.
We state the prior in Section 8.2.}

}

\subsection{Model}

\alert{We use the semiparametric population model introduced in Sections \ref{parametric} and \ref{population.model},
assuming that $I_i - E_i$ and $R_i - I_i$ are independent Gamma$(\eta_{E,1},\, \eta_{E,2})$ and Gamma$(\eta_{I,1},\, \eta_{I,2})$ random variables,
respectively ($i = 1, \dots, 186$).
The priors of the epidemiological parameters are given by $\eta_{E,1} \sim \text{Uniform}(4,\, 8)$, $\eta_{E,2} \sim \text{Uniform}(.75,\, 3)$, $\eta_{I,1} \sim \text{Uniform}(1.5,\, 8)$, $\eta_{I,2} \sim  \text{Uniform}(2.5,\, 7.5)$, and $\beta \sim \text{Uniform}(.1,\, 8)$,
which cover a range of plausible values \citep[see, e.g., the discussions of][]{GrWeHu10,GrWeHu11}.
The degree parameters $\theta_1, \dots, \theta_{186}$ are assumed to have been generated by a Dirichlet process prior with concentration parameter $\alpha$ and base distribution $N(\mu,\, \sigma^2)$. 
To express the uncertainty about the concentration parameter $\alpha$ and the parameters $\mu$ and $\sigma^2$ of the base distribution,
we assume that the hyperparameters $\alpha$, $\mu$, and $1/\sigma^{2}$ have Gamma$(5,\, 1)$,
$N(0,\, 1)$,
and Gamma$(1,\, 10)$ hyperpriors,
respectively.
To facilitate Markov chain Monte Carlo sampling from the posterior distribution,
we truncate the Dirichlet process prior at $K=3$ as described in Section \ref{truncation}.
We choose $K=3$ because we expect two subpopulations,
one corresponding to potential superspreaders and one corresponding to all other population members,
and $K=3$ is a convenient upper bound on the number of subpopulations.
We consider two specifications of the prior probabilities of who infected whom,
because the posterior is sensitive to the prior probabilities of who infected whom.
The reason is that the MERS data set does not contain direct observations of transmissions or contacts,
and the information on transmissions is therefore limited to two sources of indirect observations:
the assessments of doctors of who infected whom and the observed infectious and removal times,
as discussed in Section 8.1.
While the observed infectious and removal times help narrow down the possible sources of infections,
there may be many possible sources of infections left.
As a consequence,
it is not surprising that the posterior is sensitive to the choice of prior probabilities of who infected whom.
We demonstrate the sensitivity of the posterior to the choice of prior in Section 8.4 by using two specifications of prior probabilities:
\bi 
\item[(a)] If the doctors assessed that population member $i$ infected population member $j$,
we specify $\varphi(\mbox{$i$ infected $j$}) = 1$ and $\varphi(\mbox{$h$ infected $j$}) = 0$ for all infected population members $h \in \{1, \dots, 186\}\, \setminus\, \{i, j\}$.
If the doctors did not specify who infected $j$ and there are $M_j \in \{1, \dots, 185\}$ infected population members $h$ satisfying $I_h < E_j < R_h$,
we specify $\varphi(\mbox{$i$ infected $j$}) = 1\, /\, M_j$\, for all infected population members $i$ satisfying $I_i < E_j < R_i$ and $\varphi(\mbox{$h$ infected $j$}) = 0$ for all other infected population members $h$.
\item[(b)] If there are $M_j \in \{1, \dots, 185\}$ infected population members $h$ satisfying $I_h < E_j < R_h$,
we specify $\varphi(\mbox{$i$ infected $j$}) = 1\, /\, M_j$ for all infected population members $i$ satisfying $I_i < E_j < R_i$ and $\varphi(\mbox{$h$ infected $j$}) = 0$ for all other infected population members $h$.
\ei
}

\alert{Last, 
but not least,
there is no evidence to suggest that the incomplete-data generating process is non-ignorable,
therefore we estimate the population model under the assumption that the incomplete-data generating process is ignorable.
It is worth noting that the assumption of ignorability is a strong assumption,
but it is convenient and defensible unless there is strong evidence to the contrary.
We discuss how to deal with non-ignorable incomplete-data generating processes in Section 9.4.
}

\subsection{Computing} 

\hide{

K=1:
    user   system  elapsed 
7734.545   14.808 7767.459 
> 7767.459 # Seconds
[1] 7767.459
> 7767.459 / 60 # Minutes
[1] 129.4577

K=3:
    user   system  elapsed 
17824.30    39.13 18149.68 
> 18149.68 # Seconds
[1] 18149.68
> 18149.68 / 60 # Minutes
[1] 302.4947

K=3 without:
    user    system   elapsed 
10893.248    18.406 11038.449 
> 11038.449 # Seconds
[1] 11038.45
> 11038.449 /60 # Minutes
[1] 183.9742

}

\begin{figure}[t]
\caption{\label{traceplots}MERS data: Trace plots of the degree parameters $\gamma_1$, $\gamma_2$, and $\gamma_3$.}\s
\begin{center}
\includegraphics[width=.3\linewidth]{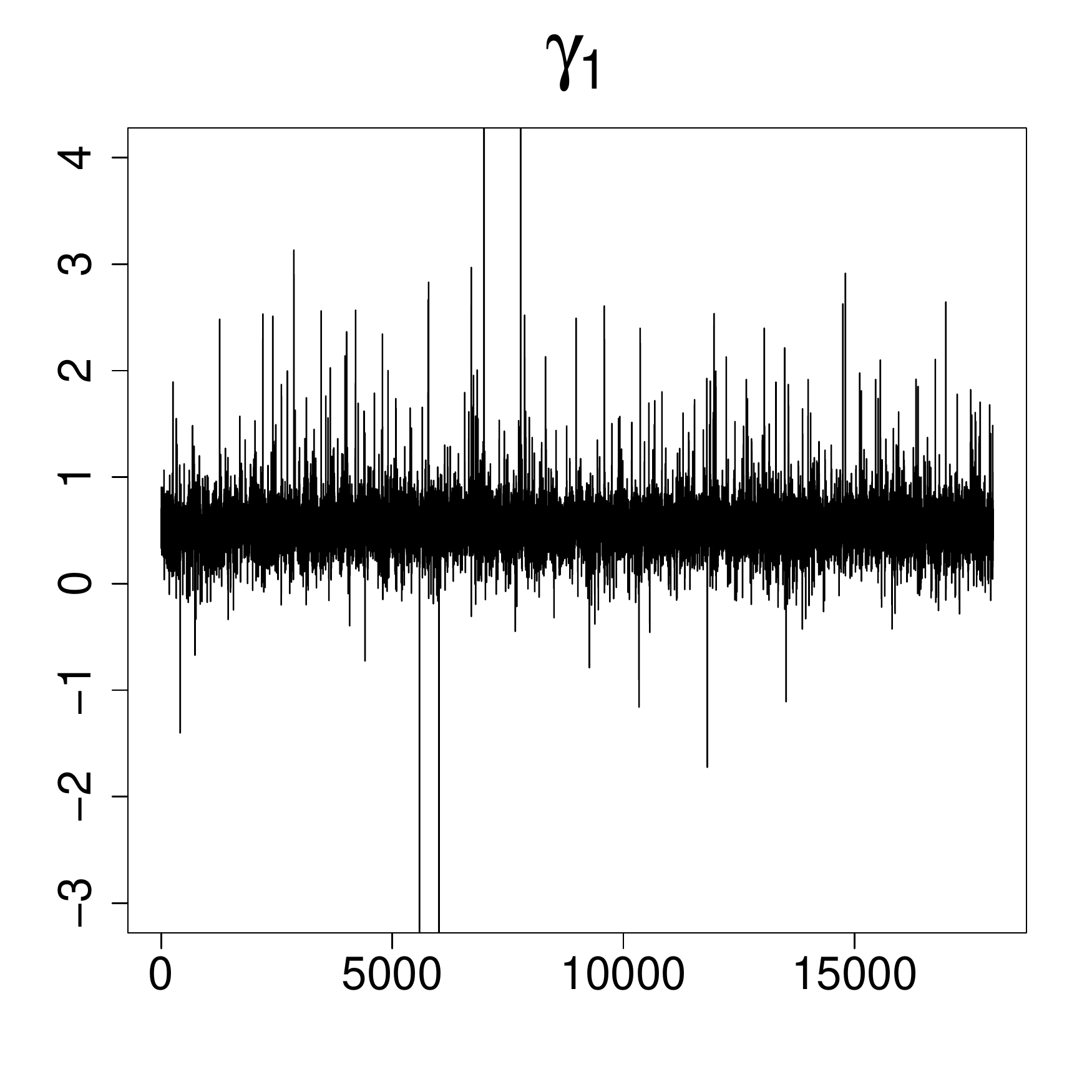}
\includegraphics[width=.3\linewidth]{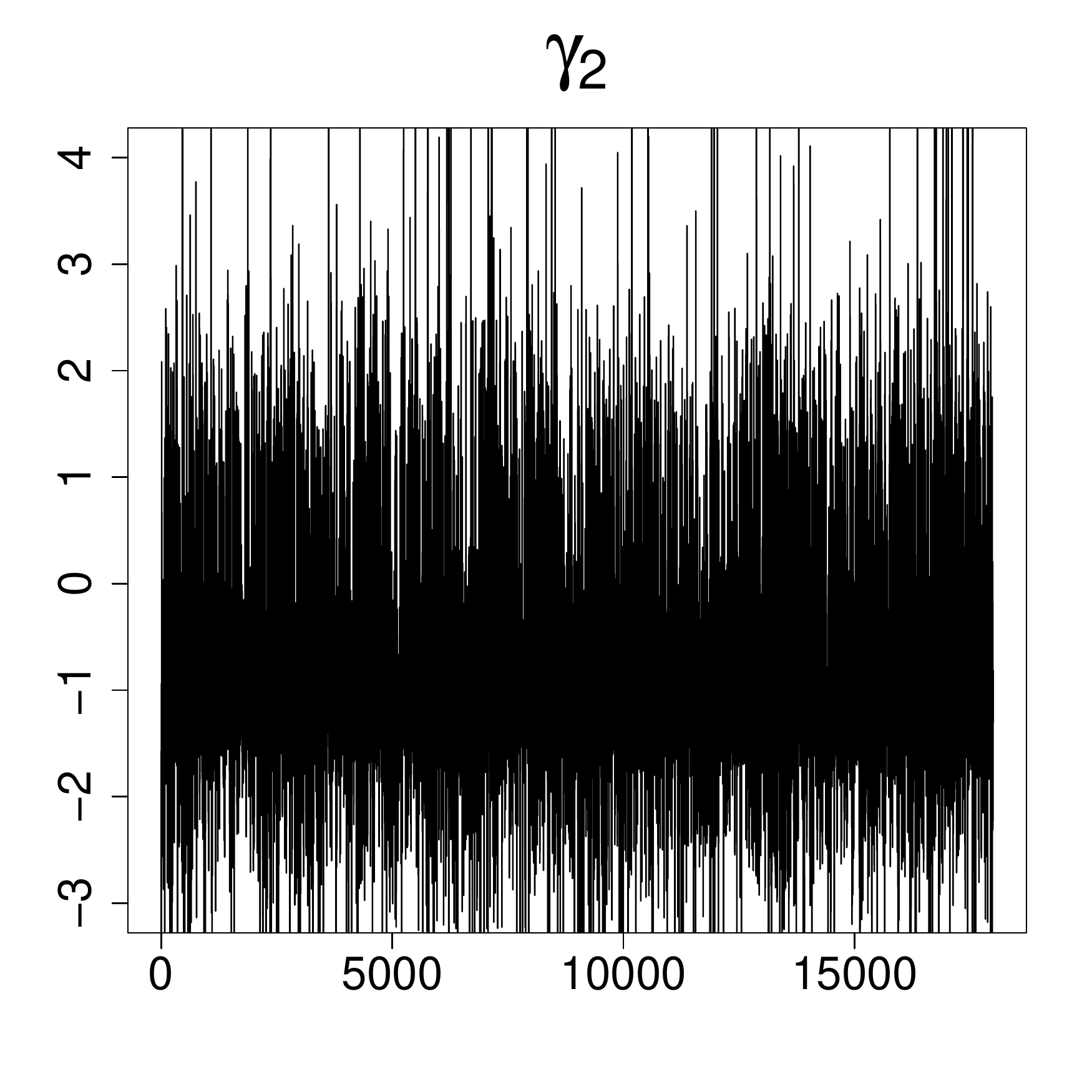}
\includegraphics[width=.3\linewidth]{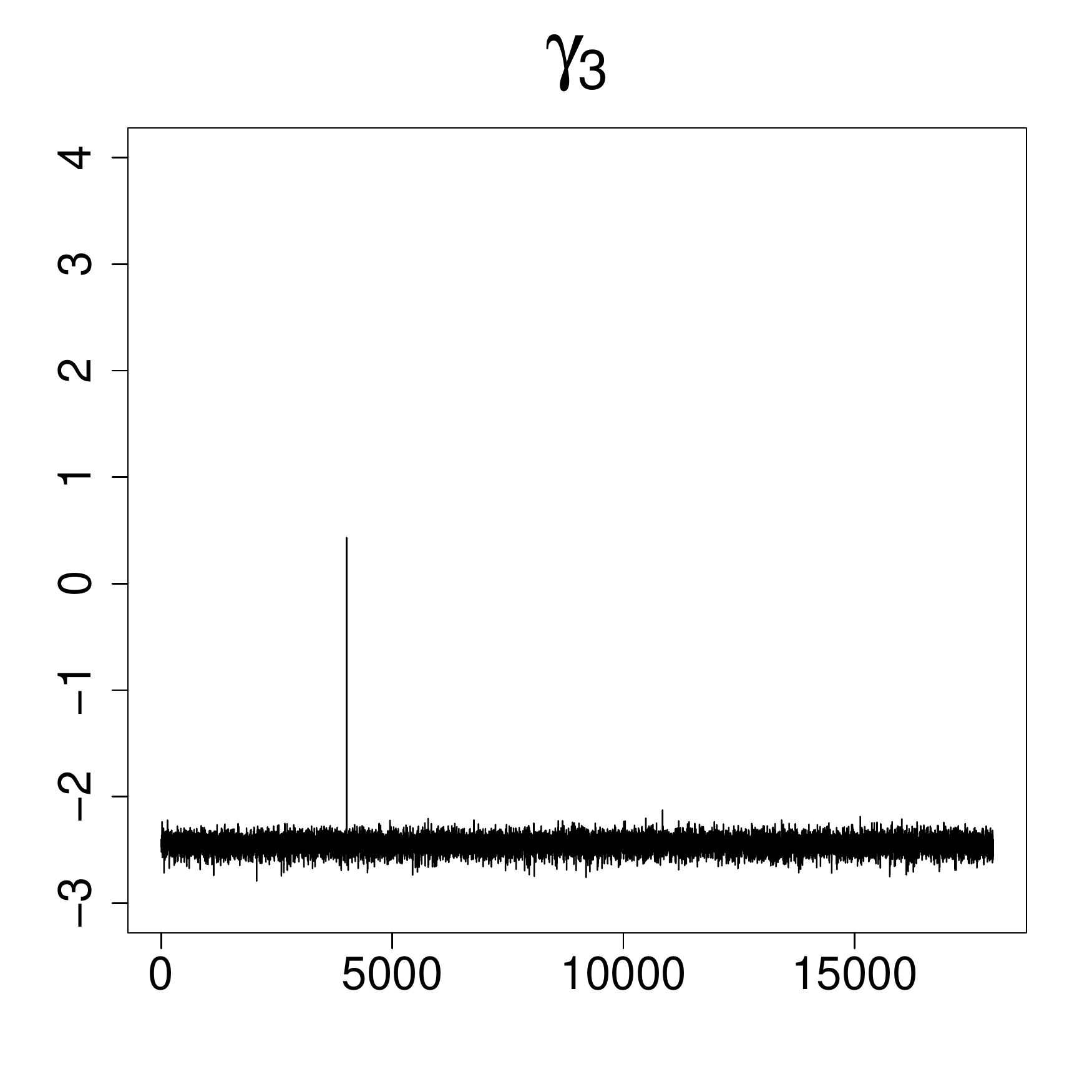}
\end{center}
\end{figure}

\alert{We sample from the posterior distribution by using the Bayesian Markov chain Monte Carlo algorithm described in Section 6.4.
The Markov chain Monte Carlo algorithm required 302 minutes.
Out of the 2,000,000 sample points generated by the Markov chain Monte Carlo algorithm,
we discard the first 200,000 sample points as a burn-in and retain every 100-th sample point of the remaining 1,800,000 sample points,
giving rise to a Markov chain Monte Carlo sample of size 18,000.
The resulting Markov chain Monte Carlo sample shows evidence of label-switching.
We undo the label-switching as described in Section 6.5.
Trace plots of selected parameters can be found in Figure \ref{traceplots}.
The trace plots do not show signs of non-convergence and suggest that the label-switching has been undone.}

\subsection{Results} 

\hide{

> table(epidata[,2])

  1   3   6   9  11  14  15  16  17  36  46  50  52  57  75  76  80  96 118 119 
 30   1   2   1   1  70   6  23   1   1   1   1   1   1   1  10   1   1   2   1 
123 132 135 143 
  1   1   2   1 

}

\hide{

> epidata[1,2]
X.1 
 NA 
> epidata[14,2]
X.1 
  1 
> epidata[16,2]
X.1 
  1 
> epidata[15,2]
X.1 
  1 
> epidata[76,2]
X.1 
 14 

}

\alert{The semiparametric population model introduced in Section 4 induces a partition of the population into subpopulations.
We therefore start by investigating the local structure of the population,
by determining which population members belong to which subpopulations.}

\begin{figure}[htp]
\caption{\label{posterior.class}
MERS data: Posterior classification probabilities of population members along with transmissions of MERS based on the assessments of doctors.
Each population member is represented by a circle.
Each circle is divided into colored slices.
The colors of the slices represent subpopulations.
The sizes of the slices are proportional to the posterior probabilities of belonging to the corresponding subpopulations.
Directed edges indicate transmissions based on the assessments of doctors.
The five population members who are believed to have infected 139 out of the 186 infected population members are labeled 1, 14, 15, 16, and 76.
The labels of all other population members are omitted.}\s
\begin{center}
\includegraphics[width=.9\linewidth]{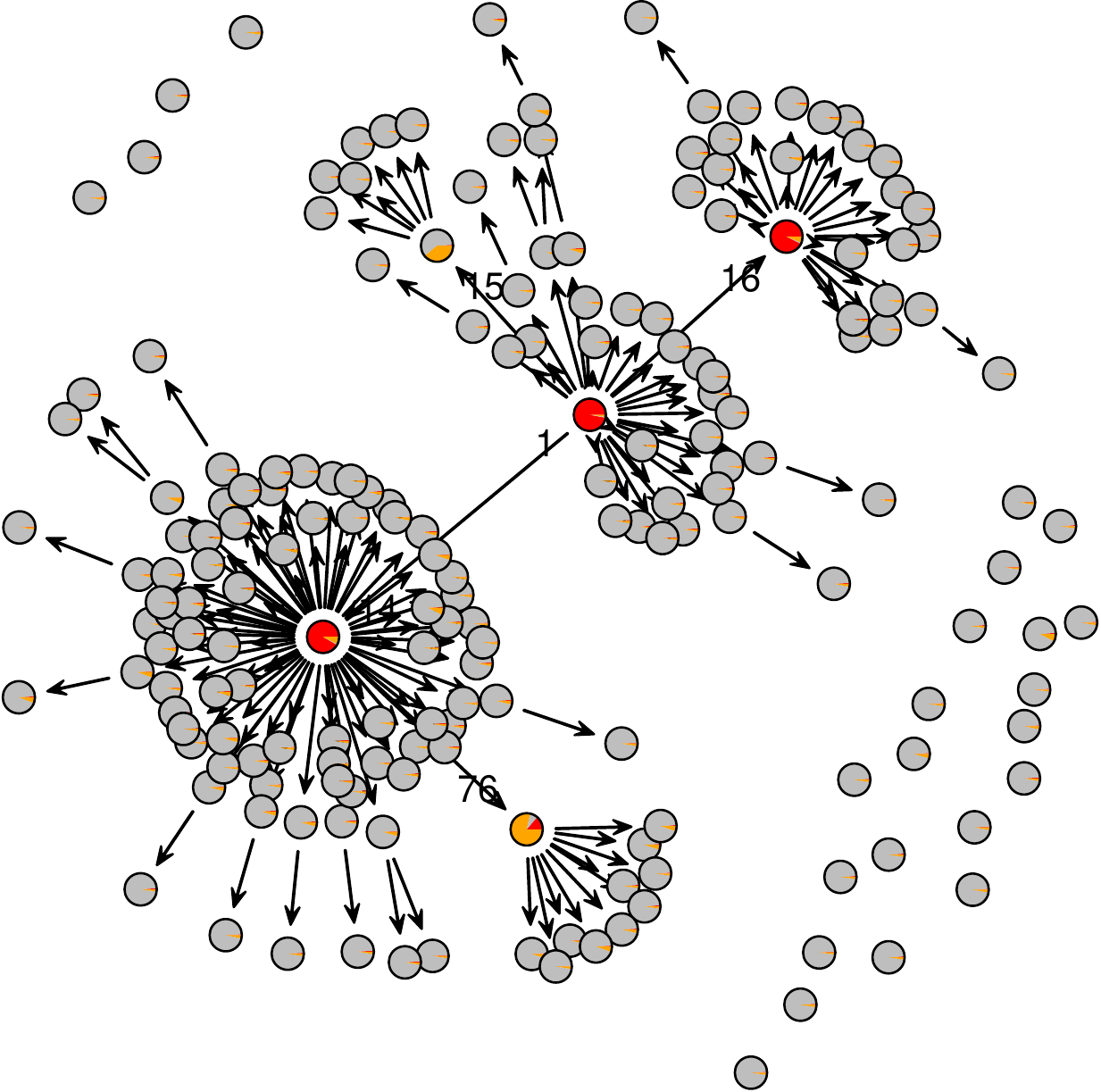}
\end{center}
\end{figure}

\alert{Figure \ref{posterior.class} shows the posterior classification probabilities of population members along with the transmissions of MERS according to the assessments of doctors.
According to doctors,
181 out of the 186 infected population members are suspected of having infected 0, 1, or 2 other population members,
while 5 population members are believed to have infected more than 2 other population members:
1 (infected 30 others); 
14 (infected 70 others); 
15 (infected 6 others); 
16 (infected 23 others); 
and 76 (infected 10 others).
In total,
these 5 population members are believed to have infected 139 out of the 186 infected population members.
The posterior classification probabilities in Figure \ref{posterior.class} suggest that three subpopulations can be distinguished:
\bi
\item A subpopulation consisting of population members 1, 14, and 16,
which we represent by the color red in Figure \ref{posterior.class}.
Population member 1 was the first confirmed MERS case in South Korea and is believed to have infected population members 14 and 16.
These three population members are suspected of being responsible for the three largest clusters of infections,
infecting a total of 123 out of the 186 infected population members,
and may therefore be considered to be primary drivers of the MERS outbreak.
\item A subpopulation consisting of population members 15 and 76,
which we represent by the color orange in Figure \ref{posterior.class}.
Population member 76 (infected 10 others) belongs with high posterior probability to the orange-colored subpopulation,
whereas there is more uncertainty about the classification of population member 15 (infected 6 others).
Both of them are thought to be responsible for clusters of infections---albeit smaller clusters of infections than population members 1, 14, and 16---and may therefore be considered to be secondary drivers of the MERS outbreak.
\item A subpopulation consisting of all other population members,
which we represent by the color gray in Figure \ref{posterior.class}.
\ei
To gain insight into the propensities of population members to form contacts within and between these subpopulations,
we inspect the marginal posterior densities of the degree parameters $\gamma_1$, $\gamma_2$, and $\gamma_3$.
The degree parameters $\gamma_1$, $\gamma_2$, and $\gamma_3$ correspond to the red-,
orange-,
and gray-colored subpopulations in Figure \ref{posterior.class},
respectively.
\begin{figure}[t]
\caption{\label{marginal.posteriors}
MERS data: Markov chain Monte Carlo approximations of the marginal posterior densities of the degree parameters $\gamma_1$, $\gamma_2$, and $\gamma_3$.
The degree parameters $\gamma_1$, $\gamma_2$, and $\gamma_3$ correspond to the red-,
orange-,
and gray-colored subpopulations in Figure \ref{posterior.class},
respectively.}\s
\begin{center}
\includegraphics[width=.3\linewidth]{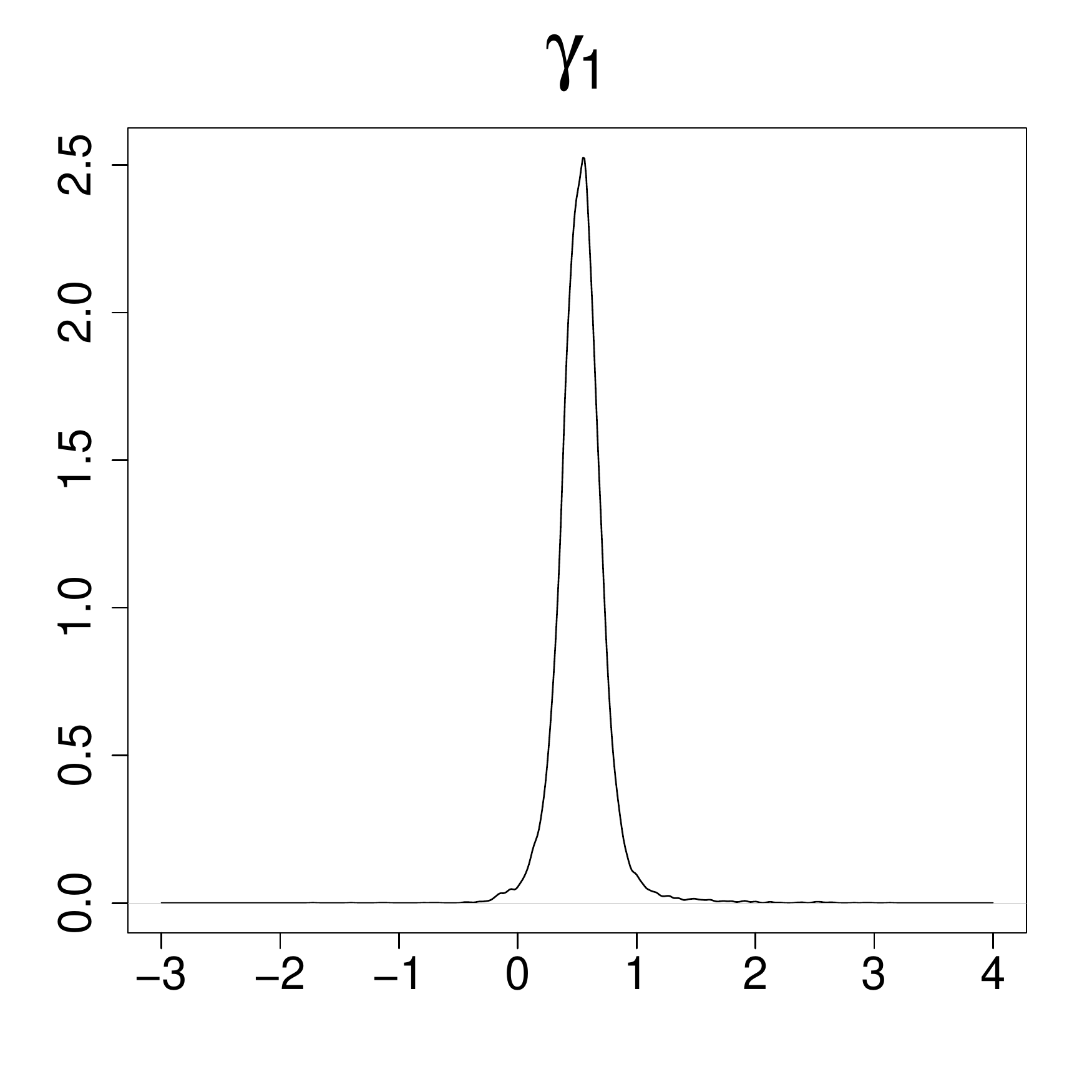}
\includegraphics[width=.3\linewidth]{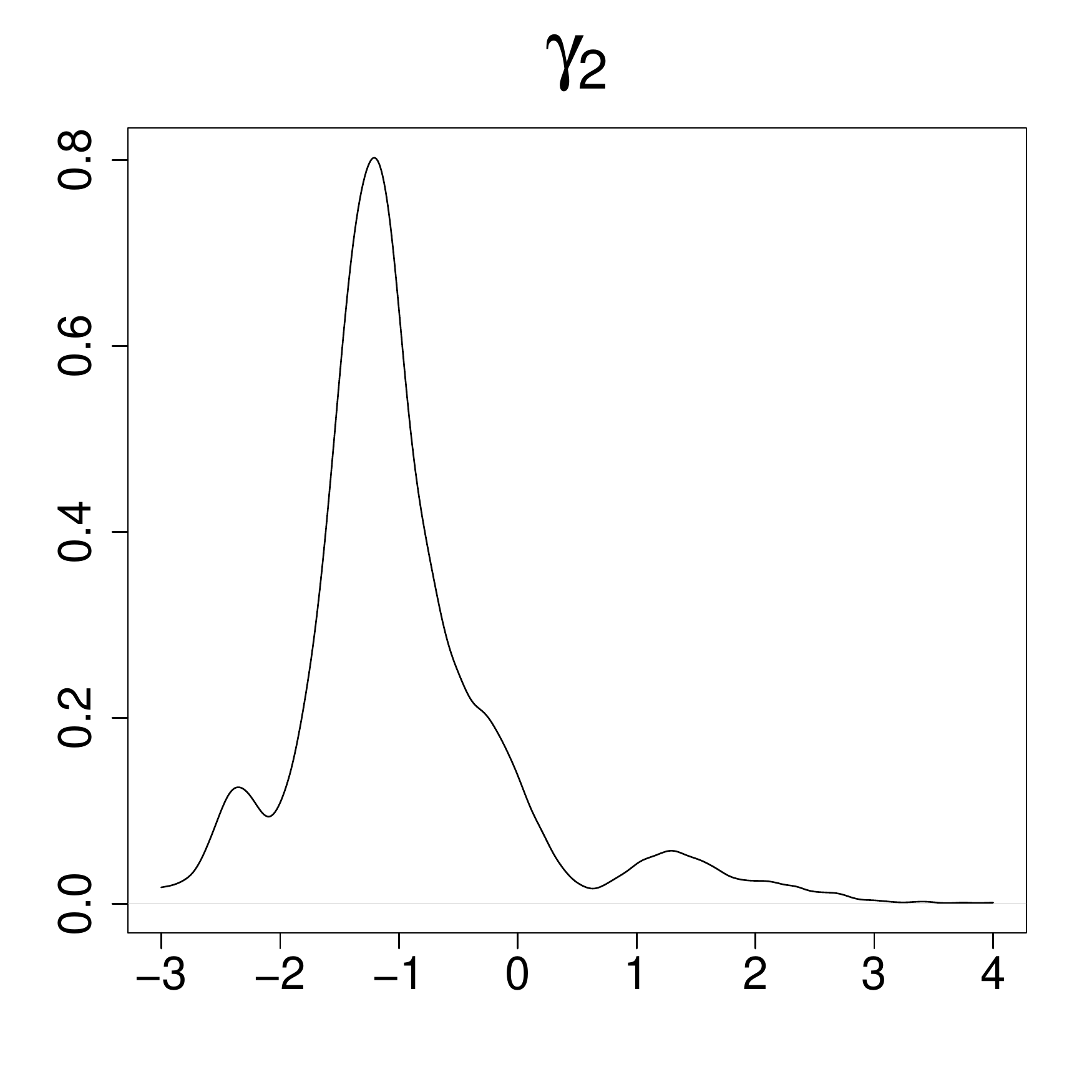}
\includegraphics[width=.3\linewidth]{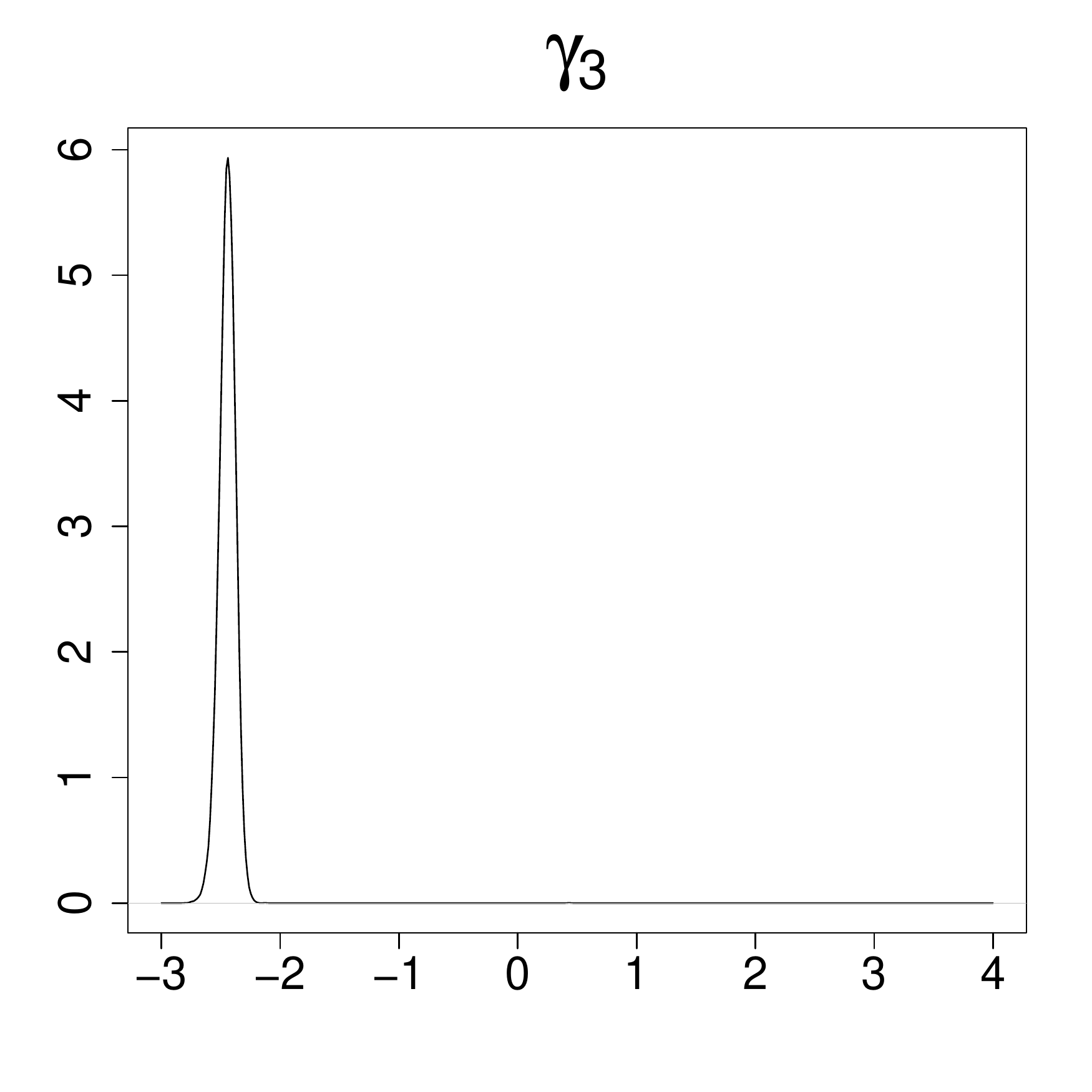}
\end{center}
\end{figure}
Figure \ref{marginal.posteriors} shows Markov chain Monte Carlo approximations of the marginal posterior densities of $\gamma_1$,
$\gamma_2$,
and $\gamma_3$.
The 95\% posterior credible intervals of $\gamma_1$,
$\gamma_2$,
and $\gamma_3$ are given by $[.140,\, .959]$,
$[-2.575,\, 1.759]$,\;
and $[-2.587,\, -2.317]$,\;
respectively.
Since the log odds of the probability of a contact between two population members $i$ and $j$ is $\theta_i + \theta_j = \bZ_i^\top\, \bgamma + \bZ_j^\top\, \bgamma$,
members of the red-colored subpopulation have a high propensity to be in contact with other population members,
in particular other members of the red- and orange-colored subpopulations,
in addition to members of the gray-colored subpopulation.
It is worth noting that the posterior uncertainty---as measured by the lengths of the 95\% posterior credible intervals---is highest for $\gamma_2$ and lowest for $\gamma_3$, 
which makes sense because no more than 1 or 2 population members belong to the orange-colored subpopulation with degree parameter $\gamma_2$ (with a high posterior probability),
whereas 181 population members belong to the gray-colored subpopulation with degree parameter $\gamma_3$ (with a high posterior probability).
Taken together,
these results suggest that there were three to five potential superspreaders who may have had a great impact on the outcome of the MERS outbreak in South Korea.
These observations underscore the importance of detecting potential superspreaders.}

\begin{figure}
\caption{\label{ppc.degrees}
MERS data: Posterior predictions of the degree distribution of the population contact network.}\s
\centering
  \includegraphics[width=.9\linewidth]{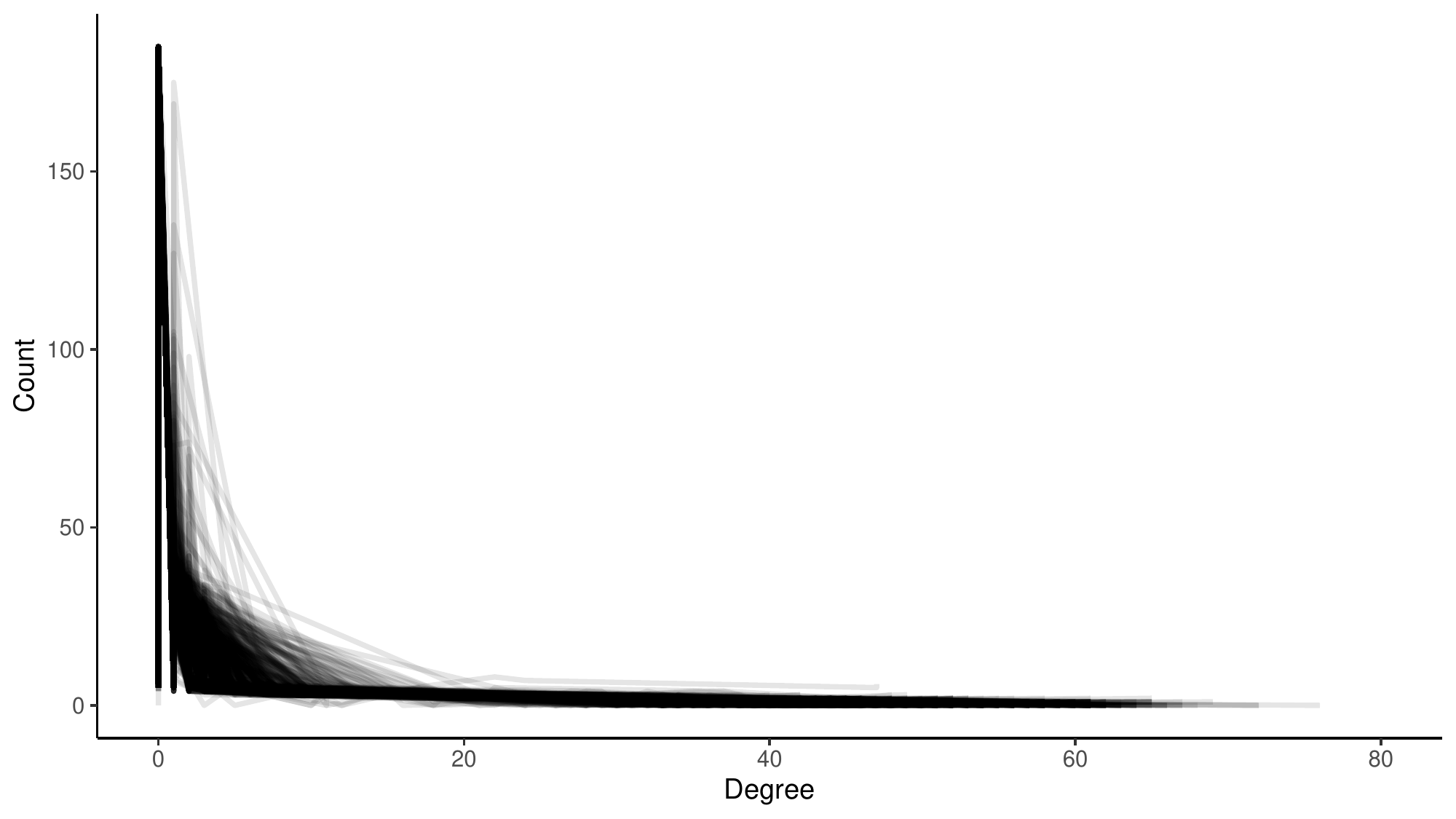}
\end{figure}

\alert{We proceed with posterior predictions of the degree distribution of the population contact network.
While the population contact network and its degree distribution are unobserved,
posterior predictions of the degree distribution can be generated,
provided draws from the posterior distribution are available.
Probabilistic statements about the degrees of population members based on the posterior distribution are informed by the observed infectious and removal times along with the assessments of doctors of who infected whom.
Both of these sources of information help inform who was in contact with whom,
because
\bi
\item the observed infectious and removal times reveal when infected population members were infectious,
which helps narrow down the possible sources of infections;
\item an infection implies a contact.
\ei
The posterior predictions of the degree distribution shown in Figure \ref{ppc.degrees} suggest that the degree distribution is long-tailed:
The bulk of population members has no more than 10 contacts,
but some population members have as many as 80 contacts.
As pointed out before,
population members with many contacts can infect many other population members and therefore represent an important public health concern.
}

\begin{figure}[htp]
\caption{\label{ppc.epidemic}
MERS data: Posterior predictions of the maximum number of infectious population members during the epidemic.
The red vertical line indicates the observed number of 129.
\hide{
The posterior predictive distribution is bimodal.
The two modes of the posterior predictive distribution correspond to two scenarios.
In the first scenario,
the first infected population member is isolated in the sense that it has no contacts.
As a result,
no other population member is infected,
giving rise to the mode at 1.
In the second scenario,
the first infected population member is not isolated.
Then MERS can spread throughout the population,
resulting in the mode at the observed number of 129.}
}
\s
\centering
\includegraphics[scale=.5]{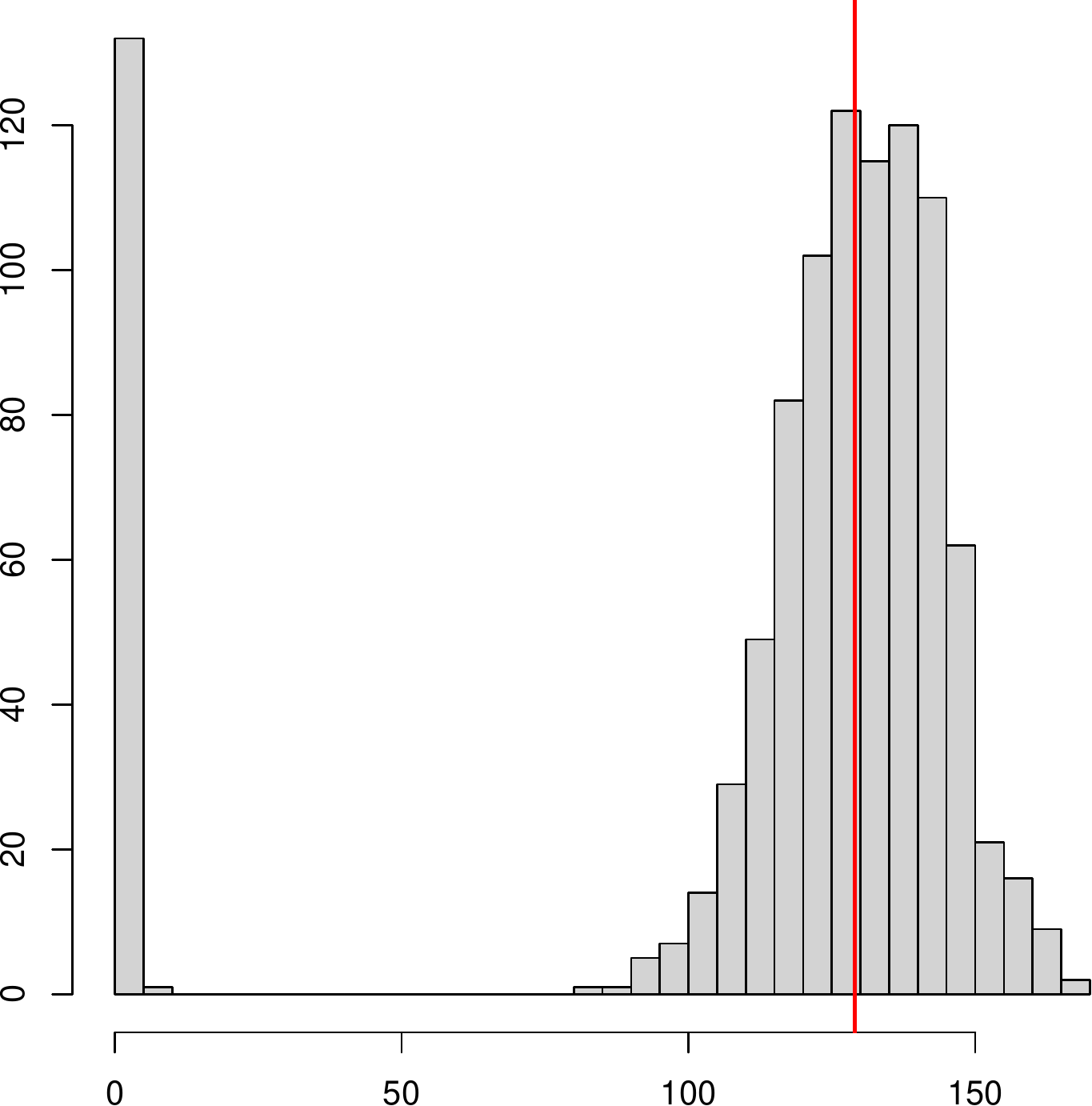}
\end{figure}

\alert{Last, 
but not least, 
it may be of interest to inspect posterior predictions of the epidemic itself.
We focus on the maximum of the ``epidemic curve,"
that is,
the maximum number of infectious population members during the height of the MERS outbreak.
The observed number is 129,
that is,
the number of infectious population members reached 129 at the height of the MERS outbreak.
Figure \ref{ppc.epidemic} reveals that the posterior predictive distribution of the maximum number of infectious population members is bimodal.
The two modes of the posterior predictive distribution correspond to two scenarios.
In the first scenario,
the first infected population member is isolated in the sense that it has no contacts.
As a result,
no other population member is infected,
giving rise to the mode at 1.
In the second scenario,
the first infected population member is not isolated.
Then MERS can spread throughout the population,
resulting in the mode at the observed number of 129.}

\alert{These results are based on prior specification (a) described in Section 8.2,
that is,
the prior probabilities of who infected whom are specified in accordance with the assessments of doctors.
If the assessments of doctors are ignored and prior specification (b) in Section 8.2 is used,
then---conditional on the event that the first infected population member is not isolated---the predicted maximum number of infectious population members is 122.5 on average, 
which is 5\% lower than the observed number of 129.
By contrast,
when the assessments of doctors are taken into account and prior specification (a) in Section 8.2 is used,
then---conditional on the event that the first infected population member is not isolated---the predicted maximum number of infectious population members is 127.9 on average,
which is 1\% lower than the observed number of 129.
In other words,
ignoring the assessments of doctors and choosing prior specification (b) leads to underpredictions of the maximum of the ``epidemic curve,"
whereas utilizing the assessments of doctors and choosing prior specification (a) leads to predictions that match,
on average,
the observed maximum of the ``epidemic curve" rather well.
These observations underscore that
\bi
\item posterior predictions of the epidemic are sensitive to the choice of prior of who infected whom,
at least in the absence of observations of transmissions or contacts;
\item data on transmissions and contacts should be collected to reduce the posterior uncertainty about quantities of interest and the sensitivity of posterior predictions of the epidemic to the choice of prior.
\ei

}

\hide{
> 122.5 / 129
[1] 0.9496124
> 127.9 / 129
[1] 0.9914729
}

\vspace{-.5cm}

\section{Open questions and directions for future research}
\label{discussion}

We have introduced a semiparametric population model that can accommodate both short- and long-tailed degree distributions and detect potential superspreaders,
in addition to dealing with a wide range of missing and sample data.

Having said that,
many open questions remain.
Some of them are rooted in the lack of data,
while others stem from computational and statistical challenges arising from the lack of data and the complexity of the models.
We review a selection of open questions and directions for future research below.

\subsection{What is the population of interest?}

To analyze the MERS outbreak in South Korea,
we applied the proposed semiparametric modeling framework to the 186 infected population members.
In so doing,
we made the implicit assumption that the population of interest consists of those 186 infected population members.
There is no denying that such an assumption is unappealing.
In fact,
the assumption was motivated by convenience rather than substantive considerations,
including the challenge of determining the population of interest:
e.g.,
does the population of interest consist of all residents of South Korea,
all residents of East Asia,
or the whole world?

\subsection{Incomplete data}
\label{inc}

As pointed out before,
collecting complete data on population contact networks and epidemics is all but impossible.
As a consequence,
public health officials and researchers face a recurring question in the event of epidemics (whether outbreaks of coronaviruses such as COVID-19, MERS, or SARS, Ebola viruses, or other viruses): 
Which data to collect?
We believe that,
to learn the structure of a population contact networks and its impact on the spread of an infectious diseases,
 investigators should attempt to collect contact and epidemiological data on all infected population members and collect samples of contacts from non-infected population members by likelihood-ignorable sampling designs.
We discuss some of the challenges arising in practice along with possible solutions.

First,
while it is challenging to collect data on transmissions,
it is advisable to collect data that help reduce the posterior uncertainty about epidemiological parameters.
One possible source of data are viral genetic sequence data,
among other possible data sources.
We refer interested readers to \citet{beast} for a recent review of possible data sources.

Second,
it is prudent to sample from population contact networks to reduce the posterior uncertainty about network parameters.
The two likelihood-ignorable sampling designs discussed above can be used to do so,
though both require a sampling frame.
If no sampling frame is available,
an alternative would be respondent-driven sampling \citep[e.g.,][]{GiHa10,Gi11},
which is a form of link-tracing without a sampling frame.
Location data collected by mobile phones and other electronic sources would be alternative sources of data,
but raise data privacy issues \citep{FiSa10}.
In the past decade,
substantial progress has been made on data privacy in the statistical literature:
see,
e.g.,
the work of \citet{karwa2016inference} on data privacy in scenarios where network data are generated by the $\beta$-model (albeit without Dirichlet process priors and without epidemics).
Studying data privacy for epidemics would be an important direction of future research.

\subsection{Computational challenges arising from incomplete data}
\label{comp}

The lack of data has computational implications.
\alert{Statistical algorithms for likelihood-based statistical inference \citep*[e.g., EM algorithms,][]{DeLaRu77} have to integrate over the unobserved data,
hence the computing time tends to increase with the amount of missing data.}
How to develop scalable statistical algorithms,
with statistical guarantees, 
is an open question.
One idea would be to develop a two-step estimation algorithm,
first estimating the network parameters and then estimating the epidemiological parameters,
leveraging computational advances in the statistical analysis of network data \citep[e.g.,][]{RaNiHoYe12,STMu13} and epidemiological data \citep{beast}.
A two-step estimation algorithm may require data on contacts,
however,
because without data on contacts the posterior correlations of the network parameters and the rate of infection can be high \citep{GrWeHu10},
in which case two-step estimation algorithms may not work well.
A related problem is how to update posteriors in reasonable time as more data on infections and contacts come in.

\subsection{Non-ignorable incomplete-data generating processes}

We have considered here ignorable incomplete-data generating processes.
If the incomplete-data generating process is non-ignorable,
then either the incomplete-data generating process must be modeled or it must be demonstrated that Bayesian inference for the parameters of the population model is insensitive to the incomplete-data generating process.
Both approaches require insight into the incomplete-data generating process and may require additional model assumptions,
some of which may be untestable.

\subsection{Population models capturing additional network features}

We have focused here on the degrees of population members as network features,
but there are many other important network features:
e.g.,
if population contact networks exhibit closure \citep[e.g., transitive closure,][]{WsFk94,Ko09a},
then infectious diseases may spread rapidly within subpopulations but may spread only slowly through the whole population,
which has potential policy implications.
There are two broad approaches to capturing closure in population contact networks:
latent space models \citep*{HpRaHm01,HaRaTa07,SmAsCa19} and related latent variable models \citep*[e.g.,][]{latentspace,RaFrRa15,FoHo15,Ho18};
and exponential-family models of random graphs \citep*{FoSd86,SnPaRoHa04,ergm.book}.
Both classes of models can accommodate the degree terms we have used \citep*[see, e.g.,][]{KrHaRaHo07,thiemichen2016bayesian} along with additional terms that reward closure in population contact networks.
\alert{However,
both of them come at costs in terms of computing time \citep*[][]{BaBrSl11,ChDi11},
although the computing time depends on the class of models under consideration \citep*[see, e.g.,][]{karwa2016dergms}.}
In addition,
\citet{We11} pointed out that closure in population contact networks may not be detectable unless data on the population contact network are collected.
This,
too,
underscores the importance of collecting network data.

\subsection{Time-evolving population contact networks}
\label{temporal}

We have assumed that the population contact network is time-invariant,
motivated by the lack of data on the population contact network and the desire to keep the model as simple and parsimonious as possible.
In practice,
the population contact network may evolve over time,
because population members may create or discontinue contacts and because authorities may enforce social distancing measures.
As a consequence,
it would be natural to allow the population contact network to change over time.
Extensions to time-evolving population contact networks could be based on temporal stochastic block and latent space models \citep*[e.g.,][]{FuSoXi10,sewell2015latent,sewell2016latent,sewell2016model};
temporal exponential-family random graph models \citep*{RoPa01,HaFuXi10,ouzienko2011decoupled,KrHa10};
continuous-time Markov processes \citep{Sn01a};
relational event models \citep{Bu08};
and other models \citep[e.g.,][]{KaPr59,DuDu14,sewell2017network}.
An interesting approach to time-evolving population contact networks was recently introduced by \citet{Buetal21},
albeit without the ability to accommodate long-tailed degree distributions and detect potential superspreaders.

\hide{

\subsection{R code}

The {\tt R} code used here is based on an extension of {\tt R} package {\tt epinet} \citep{epinet},
documented in \citet{epinet.jss}.
It can be obtained from the authors.

}

\section*{Supplementary materials}

The supplement provides a proof of Proposition \ref{theorem.2} along with details on the Bayesian Markov chain Monte Carlo algorithm.

\section*{Acknowledgements}

This work was supported by the National Institutes of Health under Grant NIH 1R01HD052887-01A2 (MS, RB);
the Office of Naval Research under Grant ONR N00014-08-1-1015 (MS);
the National Science Foundation under Grants NSF DMS-1513644 and NSF DMS-1812119 (MS, SB);
and the Army Research Office under Grant ARO W911NF-21-1-0237 (MS).

\section*{Disclosure statement}

We do not have financial interests or other interests related to the results presented here.

\section*{Data availability}

We retrieved the data from the website {\tt http://english.mw.go.kr} of the South Korean Ministry of Health and Welfare on September 22, 2015.
The website has been removed since,
but the data can be obtained from the authors.




\mbox{}

\newpage

\begin{appendix}

\begin{center}
\Large
Supplementary Materials:

A Semiparametric Bayesian Approach To Epidemics,\linebreak
with Application to the Spread of the Coronavirus MERS in South Korea in 2015

\hide{
\large
Michael Schweinberger
\hspace{1cm}
Rashmi P.\ Bomiriya
\hspace{1cm}
Sergii Babkin
}
\end{center}

\maketitle

The supplement is divided into the following two appendices:\s

\noindent
{Appendix \ref{appendix.b}: Proof of Proposition \ref{theorem.2}}\dotfill\pageref{appendix.b}\s
\\
{Appendix \ref{mcmc}: Bayesian Markov chain Monte Carlo algorithm}\dotfill\pageref{mcmc}

\section{Proof of Proposition \ref{theorem.2}}
\label{appendix.b} 

We prove Proposition \ref{theorem.2} stated in Section \ref{bayesian.inference} of the manuscript.\s

{\bf Proof of Proposition \ref{theorem.2}.}
Since the incomplete-data generating process is likelihood-ignorable,
\bea
\nonumber
p(\bvartheta,\, \bta,\, \btheta \mid \baa,\, \bx_{\mbox{\tiny obs}},\, \by_{\mbox{\tiny obs}})
\;=\; \dis\sum_{\by_{\mbox{\tiny mis}}}\, \dis\int p(\bx_{\mbox{\tiny mis}},\, \by_{\mbox{\tiny mis}},\, \bvartheta,\, \bta,\, \btheta \mid \baa,\, \bx_{\mbox{\tiny obs}},\, \by_{\mbox{\tiny obs}}) \dd \bx_{\mbox{\tiny mis}}\s
\\
\hspace{1cm}\;\propto\; \dis\sum_{\by_{\mbox{\tiny mis}}}\, \dis\int p(\baa,\, \bx,\, \by,\, \bvartheta,\, \bta,\, \btheta) \dd \bx_{\mbox{\tiny mis}}\s
\\
\hspace{1cm}\propto\; \dis\sum_{\by_{\mbox{\tiny mis}}}\, \dis\int p(\baa \mid \bx,\, \by,\, \bvartheta)\; p(\bvartheta)\; p(\bx \mid \by,\, \bta)\, p(\by \mid \btheta)\, p(\bta,\, \btheta)\dd \bx_{\mbox{\tiny mis}}\s
\\
\hspace{1cm}\propto\; \dis\sum_{\by_{\mbox{\tiny mis}}}\, \dis\int p(\baa \mid \bx_{\mbox{\tiny obs}},\, \by_{\mbox{\tiny obs}},\, \bvartheta)\; p(\bvartheta)\; p(\bx \mid \by,\, \bta)\, p(\by \mid \btheta)\, p(\bta,\, \btheta) \dd \bx_{\mbox{\tiny mis}}\s
\\
\hspace{1cm}\propto\; p(\baa \mid \bx_{\mbox{\tiny obs}},\, \by_{\mbox{\tiny obs}},\, \bvartheta)\; p(\bvartheta)\; \dis\sum_{\by_{\mbox{\tiny mis}}} \dis\int p(\bx \mid \by,\, \bta)\, p(\by \mid \btheta)\, p(\bta,\, \btheta) \dd \bx_{\mbox{\tiny mis}}\s
\\
\hspace{1cm}\propto\; p(\bvartheta \mid \baa,\, \bx_{\mbox{\tiny obs}},\, \by_{\mbox{\tiny obs}})\; p(\bta,\, \btheta \mid \baa,\, \bx_{\mbox{\tiny obs}},\, \by_{\mbox{\tiny obs}}),
\eea
where
\bea
\nonumber
p(\bvartheta \mid \baa,\, \bx_{\mbox{\tiny obs}},\, \by_{\mbox{\tiny obs}})
\;=\; \dfrac{p(\baa \mid \bx_{\mbox{\tiny obs}},\, \by_{\mbox{\tiny obs}},\, \bvartheta)\, p(\bvartheta)}{\int p(\baa \mid \bx_{\mbox{\tiny obs}},\, \by_{\mbox{\tiny obs}},\, \bvartheta)\, p(\bvartheta) \dd \bvartheta}\s
\\
p(\bta,\, \btheta \mid \baa,\, \bx_{\mbox{\tiny obs}},\, \by_{\mbox{\tiny obs}}) = \dfrac{\sum_{\by_{\mbox{\tiny mis}}} \int p(\bx \mid \by,\, \bta)\, p(\by \mid \btheta)\, p(\bta,\, \btheta) \dd \bx_{\mbox{\tiny mis}}}{\sum_{\by_{\mbox{\tiny mis}}} \int\int\int p(\bx \mid \by,\, \bta)\, p(\by \mid \btheta)\, p(\bta,\, \btheta) \dd \bx_{\mbox{\tiny mis}} \dd \bta \dd \btheta},
\eea
which implies that the parameter $\bvartheta$ of the incomplete-data generating process and the parameters $\bta$ and $\btheta$ of the population model are independent under the posterior.

\section{Bayesian Markov chain Monte Carlo algorithm}
\label{mcmc}

We sample from the posterior distribution by combining the following Markov chain Monte Carlo steps by means of cycling or mixing \citep*{Tl94,Li08}.

\s

{\bf Concentration parameter $\alpha \mid \bpi$.}
Assuming the hyperprior of concentration parameter $\alpha$ is Gamma$(A_1,\, B_1)$,
we can update $\alpha$ by sampling
\bea
\nonumber
\alpha \mid \bpi
&\sim& \mbox{Gamma}(A_1 + K - 1,\; B_1 - \log \pi_K).
\eea

{\bf Mean parameter $\mu \mid \sigma^2,\, \bgamma$.}
Assuming the hyperprior of mean parameter $\mu$ is $N(O,\, S^2)$,
we can update $\mu$ by sampling
\bea
\nonumber
\mu \mid \sigma^2,\, \bgamma
&\sim& N\left(\dis\frac{S^{-2}\, O + \sigma^{-2} \sum_{k = 1}^K \gamma_k}{S^{-2} + K \sigma^{-2}},\; \dis\frac{1}{S^{-2} + K \sigma^{-2}}\right).
\eea

{\bf Precision parameter $1/\sigma^{2} \mid \mu,\, \bgamma$.}
Assuming the hyperprior of precision parameter $1/\sigma^{2}$ is Gamma$(A_2,\, B_2)$,
we can update $1/\sigma^{2}$ by sampling
\bea
\nonumber
1/\sigma^2 \mid \mu,\, \bgamma
&\sim& \mbox{Gamma}\left(A_2 + \dfrac{K}2,\; B_2 + \dsum\limits_{k = 1}^K \dis\frac{(\gamma_k - \mu)^2}2\right).
\eea

{\bf Mixing proportions $\bpi \mid \bZ=\bz,\, \alpha$.}
Let $N_k$ be the number of population members in subpopulation $k$.
Then $\bpi$ can be updated by sampling
\bea
\nonumber
V_k \mid \bZ=\bz,\, \alpha
&\ind& \mbox{Beta}\left(1 + N_k,\; \alpha + \dsum\limits_{j = k + 1}^K N_j\right),\; k = 1, \dots, K - 1,
\eea
then setting $V_K = 1$ and constructing $\bpi$ as follows:
\bea
\nonumber
  \pi_1 \= V_1\s
  \\
  \pi_k \= V_k\, \dis\prod_{j = 1}^{k - 1}\, (1 - V_j),\; k = 2, \dots, K.
\eea

\s

{\bf Indicators $\bZ_i \mid \bY=\by,\, \bZ_{-i} = \bz_{-i},\, \bgamma,\, \bpi$.}
We update indicators $\bZ_i$ by sampling
\bea
\nonumber
\bZ_i \mid \bY=\by,\, \bZ_{-i} = \bz_{-i},\, \bgamma,\, \bpi
&\sim& \mbox{Multinomial}(1; \pi_{i,1}, \dots, \pi_{i,K}),
\eea
where
\bea
\nonumber
\pi_{i,k} = \dis\frac{\dis \pi_k\, \prod_{j:\, j \neq i}^N p(y_{i,j} \mid Z_{ik} = 1,\, \bY=\by,\, \bZ_{-i}=\bz_{-i},\, \bgamma,\, \bpi)}{\dis\sum_{l = 1}^K \left(\pi_l\, \prod_{j:\, j \neq i}^N p(y_{i,j} \mid Z_{il} = 1,\, \bY=\by,\, \bZ_{-i}=\bz_{-i},\, \bgamma,\, \bpi)\right)}.
\eea

\s

{\bf Degree parameters $\bgamma \mid \bY=\by,\, \bZ=\bz$.}
We update $\bgamma$ by Metropolis-Hastings steps,
where proposals are generated from random-walk, independence, or autoregressive proposal distributions \citep{Tl94}.

\s

{\bf Degree parameters $\btheta$.}
The degree parameter $\theta_i$ of population member $i$ is updated by setting
\bea
\nonumber
\theta_i
\= \bZ_i^\top\, \bgamma,
&& i = 1, \dots, N.
\eea

\s

{\bf Epidemiological parameters $\bta \mid \bX = \bx$.}
We use Gibbs and Metropolis-Hastings steps for updating $\bta$ along the lines of \citet{GrWeHu10,GrWeHu11}.
\s

{\bf Unobserved contacts $Y_{i,j} \mid \bX=\bx,\, \bZ_i=\bz_i,\, \bZ_j=\bz_j,\, \bgamma,\, \bta$.}
If $Y_{i,j}$ is unobserved,
we can update $Y_{i,j}$ by sampling
\bea
Y_{i,j} \mid \bX=\bx,\, \bZ_i=\bz_i,\, \bZ_j=\bz_j,\, \bgamma,\, \bta
&\ind& \mbox{Bernoulli}(p_{i,j}),
\eea
where
\bea
p_{i,j}
\= \dfrac{\exp(-\beta\, \max(\min(E_j, R_i) - I_i, 0))\, \mbP_{\theta_i,\theta_j}(Y_{i,j}=1)}{p_{i,j}(0) + \exp(-\beta\, \max(\min(E_j, R_i) - I_i, 0))\, \mbP_{\theta_i,\theta_j}(Y_{i,j}=1)}.
\eea
Here,
$\mbP_{\theta_i,\theta_j}(Y_{i,j}=y_{i,j})$ is defined by
\bea
\mbP_{\theta_i,\theta_j}(Y_{i,j}=y_{i,j}) \= \exp(\lambda_{i,j}(\btheta)\, y_{i,j} - \psi_{i,j}(\btheta)),
\ee
where
\bea
\lambda_{i,j}(\btheta)
&=& \theta_i + \theta_j.
\eea

\s

{\bf Unobserved transmissions $\bT \mid \bE,\, \bI,\, \bR,\, \bY = \by$.}
If population member $j$ was infected and the source of infection is unknown,
we can update the source of infection by sampling from its full conditional distribution.
If $\varphi(\mbox{$i$ infected $j$})$ denotes the prior probability of the event that population member $i$ infected population member $j$,
the conditional probability of the event that population member $i$ infected population member $j$,
given everything else,
takes the form
\bea
\nonumber
\mbP(\mbox{$i$ infected $j$} \mid \bE,\, \bI,\, \bR,\, \bY=\by)
&=& \dfrac{y_{i,j}\; 1_{I_i < E_j < R_i}\; \varphi(\mbox{$i$ infected $j$})}{\sum_{h=1:\, h \neq j}^M\, y_{h,j}\; 1_{I_h < E_j < R_h}\; \varphi(\mbox{$h$ infected $j$})}.
\eea

\s

{\bf Unobserved exposure, infectious, and removal times $\bE,\, \bI,\, \bR \mid \bT,\, \bta$.}
If some or all exposure and infectious times $\bE$ and $\bI$ are unobserved,
we use the Metropolis-Hastings steps of \citet{GrWeHu10} for updating them.
If some or all removal times $\bR$ are unobserved,
we use the Gibbs and Metropolis-Hastings steps described in the Ph.D.\ thesis of \citet[][Section 4.6]{Bo14} for updating them.

\end{appendix}

\end{document}